\documentclass[fleqn,usenatbib]{mnras}

\usepackage{newtxtext,newtxmath}
\usepackage[T1]{fontenc}

\DeclareRobustCommand{\VAN}[3]{#2}
\let\VANthebibliography\thebibliography
\def\thebibliography{\DeclareRobustCommand{\VAN}[3]{##3}\VANthebibliography}

\usepackage{graphicx}	% Including figure files
\usepackage{amsmath}	% Advanced maths commands
\usepackage{amssymb}	% Extra maths symbols
\usepackage{lipsum}     % Placer text
\usepackage{siunitx}    % \num command for scientific notation
\usepackage{bm}         % bm for bold math
\usepackage{orcidlink}

\title[``Old age should burn and rave'']{Chaotic winds from a dying world:\\a one-dimensional map for evolving atmospheres}
\author[Bromley \& Chiang]{Joshua Bromley$^{1,2}$\thanks{email: jbromley@berkeley.edu}\orcidlink{},
Eugene Chiang$^{1,3}\thanks{email: echiang@astro.berkeley.edu}\orcidlink{0000-0002-6246-2310}$
\\
$^{1}$Department of Astronomy, University of California, Berkeley, Berkeley, CA 94720-3411\\
$^{2}$Department of Physics, University of California, Berkeley, Berkeley, CA 94720-7300\\
$^{3}$Department of Earth and Planetary Science, University of California, Berkeley, Berkeley, CA 94720-4767\\
}
\pubyear{2023}

\begin{document}
\label{firstpage}
\pagerange{\pageref{firstpage}--\pageref{lastpage}}
\maketitle

\begin{abstract}
Planets which are smaller than Mercury and heated to sublimation temperatures of $\sim$2000 K lose mass catastrophically in dusty evaporative winds. The winds are observed to gust and recede largely without pattern; 
%as indicated by 
transit depths from the {\it Kepler} mission
%that 
vary randomly from orbit to orbit by up to a factor of 10 or more.
%wildly, as indicated by large and random orbit-to-orbit transit depth variations from the {\it Kepler} mission. %observed by the {\it Kepler} transit mission, the final death throes are convulsive, with outflows of gas and dust occurring at random intervals and with periods of calm broken by outbursts of gas and dust. 
We explain how chaotic outflows may arise by constructing a map for the wind mass-loss rate as a function of time. The map is built on three statements: (1) The wind mass-loss rate scales in proportion to the surface equilibrium vapor pressure, rising exponentially with ground temperature. (2) Because the wind takes a finite time to escape the planet's gravity well, the surface mass-loss rate at any time determines the wind optical depth at a later time---the atmosphere has hysteresis. %this introduces a feedback delay between the ground and the wind. 
(3) The ground temperature increases with optical depth (greenhouse effect) when the atmosphere is optically thin, and decreases with optical depth when the atmosphere is optically thick (nuclear winter). Statement (3) follows from how dust condenses in the face of intense stellar irradiation. As discussed recently, condensates initially naked before the star must be silicate-rich and iron-poor, staying cool enough for condensation by absorbing weakly in the visible and emitting strongly in the infrared. Later, when grains are numerous enough to self-shield from starlight, they may accrete more iron and reverse their visible-to-infrared opacity ratio. Depending on parameters, the map for the wind can regularly boom and bust between a greenhouse and a nuclear winter, or erupt into chaos. Lyapunov times are measured in orbital periods, the time for the wind to turn by Coriolis forces away from the planet's dayside, out of the Hill sphere.
\end{abstract}


%%%%%%%%%%%%%%%%%%%%%%%%%%%%%%%%%%%%%%%%%%%%%%%%%%

%%%%%%%%%%%%%%%%% BODY OF PAPER %%%%%%%%%%%%%%%%%%

\section{Introduction}\label{sec:intro}
In its search for life-bearing planets, the {\it Kepler} spacecraft has discovered worlds that are dying---vaporizing under the glare of their host stars. The class of disintegrating planets include the archetype KIC 12557548b (hereafter KIC 1255b; a.k.a.~Kepler-1520b; \citealt{rappaport12}), KOI-2700b (\citealt{rappaport14}), and K2-22b (\citealt{sanchisojeda15}). These planets orbit late-type stars with periods on the order of $\sim$10 hours and have effective temperatures in excess of $\sim$2000 K, hot enough to sublimate silicates and irons off their surfaces. That mass is being lost from these bodies is indicated by the `shark-tooth' shapes of their transit light curves, signifying comet-like tails filled with particulates (\citealt{lecavelierdesetangs99}; \citealt{rappaport12}; \citealt{brogi12}; \citealt{vanlieshout16}; \citealt{schlawin21}, and references therein). Disintegrating rocky planets must have masses less than about Mercury's ($\lesssim 0.1 M_\oplus$) to ensure that their surface gravities are low enough to permit evaporative winds to escape to infinity (\citealt{perezbecker13}). The winds tend to super-saturate as they expand and cool, allowing dust to condense out of the flow (\citealt{booth23}) and obscure the host star, as observed.

A planet that perishes from thermal vaporization does not go gentle into that good night. Because the wind mass-loss rate is exponentially sensitive to surface gravity, a planet that begins its life on a 10-hr orbit with Mercury's mass can simmer for Gyrs before erupting in an outpouring of gas lasting only dozens of Myrs (\citealt{perezbecker13}). The planet KIC 1255b, estimated to have a present-day mass comparable to the Moon's ($0.01 M_\oplus$), has likely entered this final catastrophic phase, as have KOI-2700b and K2-22b, all of which sport dusty comae and tails covering up to tens of thousands of times more occulting area than their underlying planets' hard sphere surfaces.

The end of life also appears wracked by convulsions. Transit depths for KIC 1255b can vary by up to a factor of 10 or more from transit to transit (\citealt{rappaport12}; \citealt{vanwerkhoven14}). Similar fluctuations are seen at less signal-to-noise for the other systems. The transit depth variations imply that the cometary tails wax and wane on orbital timescales, and seemingly in random fashion. In a couple of instances KIC 1255b exhibited an `on-off' pattern where a deep $> 0.5\%$ transit alternated with no transit signal, for a duration of $\sim$10 orbits (\citealt{vanwerkhoven14}). In another two episodes, the system was `off' for several dozens of consecutive orbits (\citealt{rappaport12}; \citealt{kawahara13}; \citealt{vanwerkhoven14}; \citealt{croll15}; \citealt{schlawin18}). 

One mechanism for time variability, outlined by \cite{rappaport12}, involves a limit cycle that switches between low and high outflow rates. %(Rappaport et al.~2012). 
Under `clear skies', light from the host star heats the planetary surface and drives a strong evaporative wind. The wind is self-limiting, as dust that condenses from the wind attenuates incoming starlight and cools the surface. Once skies become too `cloudy', the evaporative wind shuts off; during this quiescent period, the atmosphere clears, and the cycle restarts. \cite{perezbecker13} reasoned that the time between on and off phases---the atmospheric `refresh time'---would be the time for the wind to travel from its substellar launch point to the planet's Hill sphere radius. By then, the Coriolis force would have turned the wind by an order-unity angle,  and the flow at such distance would no longer intercept stellar radiation directed at the planet's substellar point.

The wind travel time to the Hill sphere boundary is comparable to the orbital period. By definition, on Hill sphere scales, the tidal gravity from the star acting to accelerate material off the planet is competitive with the planet's gravity. Not surprisingly then, 
%the planet's wind is accelerated through the Mach 1 sonic point near the Hill radius (\citealt{perezbecker13}; Murray-Clay et al.~2009). For a trans-sonic wind, the sonic point is a critical point where the local sound speed nearly matches the local escape velocity (Parker 1958; see also the textbook by Frank, King, \& Raine 2000). 
the wind speed near the Hill radius is comparable to the local escape velocity---this is confirmed in spatially resolved models for a steady trans-sonic Parker wind from a planet in the stellar tidal potential (\citealt{murrayclay09}; \citealt{perezbecker13}). Thus for Hill radius $R_{\rm H} \sim (M/M_\star)^{1/3}a$, orbital radius $a$, planet mass $M$, host star mass $M_\star$, and gravitational constant $G$, the wind travel time is given by
\begin{equation}
t_{\rm travel} \sim \frac{R_{\rm H}}{\sqrt{GM/R_{\rm H}}} \sim \frac{a^{3/2}}{\sqrt{GM_\star}} 
\end{equation}
i.e. of order the orbital period. Over this time, the Coriolis force produces a velocity $\sim$$\Omega u \cdot t_{\rm travel} \sim u$, perpendicular to the original flow velocity $u$, for orbital frequency $\Omega = \sqrt{GM_\star/a^3}$. A more precise and spatially resolved calculation by \cite{perezbecker13} that solves explicitly for how a steady wind starts subsonically from the planet's surface, and accelerates to the Mach = 1 sonic point just shy of the Hill radius,\footnote{For a trans-sonic wind, the sonic point is a critical point where the local sound speed nearly equals the local escape velocity (\citealt{parker58}; see, e.g., chapter 2 of  \citealt{frank02}).} yields a surface-to-Hill-sphere travel time of 13 hr for KIC 1255b, nearly equal to its orbital period of 15.7 hr.

The not coincidental near-match between the orbital period and the atmospheric refresh time (as given by $t_{\rm travel}$) explains why the transit depth can vary from transit to transit in the proposed limit cycle. Beyond this sketch, however, not much progress in our understanding of time variability seems to have been made. Why should a self-limiting wind not relax to a unique equilibrium instead of cycling between a high state and a low state? Moreover, a 2-cycle is not chaotic, whereas the observations show that transit depths are stochastic, falling into an on-off pattern only occasionally and briefly. Some of the radiative-hydrodynamic wind models of \cite{booth23} exhibit cycling behaviour, but on timescales of $\sim$$10^3$ s, a factor of 40 shorter than the orbital period. This short timescale arises in their models from the assumed cm-scale thickness of the thermal boundary layer just underneath the planet's surface. The amplitude of the modeled variability is much reduced when averaged over the atmospheric refresh time of $\sim$10 hr, and may not be sufficient to reproduce the factor of > 10 variations in transit depth observed.

In this paper we seek to develop a better understanding of the time-variable death throes of vaporizing rocky planets. Rather than continue to pursue complicated and costly  radiation-chemical-hydrodynamic simulations  (\citealt{perezbecker13}; \citealt{kang21}; \citealt{booth23}), we employ a simple, well-worn tool in nonlinear dynamics, the discrete one-dimensional map. Low-dimensional maps have been used to study the regular and chaotic dynamics of plasmas and particle accelerators (\citealt{chirikov71}), planetary $N$-body systems (\citealt{wisdom82, wisdom83} ; \citealt{wisdom91}; \citealt{duncan89}), atmospheric convection (\citealt{henon76}), and biological populations (\citealt{may76}), to name just a few applications (see \citealt{strogatz15} for an introduction). Here we develop a map for how the mass-loss rate and optical depth of a thermal planetary wind evolve in time. We strip the problem down to relate just three quantities: the planet surface temperature, the wind mass-loss rate, and the optical depth between the star and the planet surface. Our goal is to construct a minimalist model for the time variability exhibited by disintegrating planets, sacrificing realism for insight into the ingredients for limit cycles and/or chaos. Hopefully our bare-bones map can guide and inspire future calculations that are better resolved in time and space and that better reproduce the observations. While we do not model the cometary tail and indeed do not explicitly resolve any flow property in any direction, the wind's total optical depth, which we follow at both stellar (visible) and reprocessed (infrared) wavelengths, may be considered a proxy for transit depth. Accordingly we will focus on how optical depth evolves with time in our model.

Section \ref{sec:theory} constructs and explores three maps. Section \ref{sec:sum} summarizes and discusses.

\section{One-Dimensional Maps}\label{sec:theory}
We construct a model
for the time evolution of
the mass-loss rate $\dot{M}$
from an evaporating planet.
The model is not resolved
in space but is resolved
in time on a discrete
grid: it maps
the mass-loss rate $\dot{M}(i)$ at
time $i$ to the mass-loss
rate at the next time  $\dot{M}(i+1)$.

We begin by positing
that 
$\dot{M}$ at any time is exponentially
sensitive to the dayside surface
temperature $T$ at that
same time:
\begin{align}
\dot{M}(i) = c_1 \exp [-c_2/T(i)] \label{mdot}
\end{align}
for positive constants
$c_1$ and $c_2$. The form
of equation (\ref{mdot})
is inspired by
the Clausius-Clapeyron
equation for the 
vapor pressure above
a solid.

We next declare that the surface temperature
$T$ depends on the planetary atmosphere's infrared optical
depth $\tau$ as
\begin{align}
T(i) &= c_3 [  (1+1/\gamma) + (1-1/\gamma)e^{-\gamma\tau(i)}]^{1/4} \label{Tg}
\end{align}
for positive constant $c_3$. 
The form of equation (\ref{Tg}) is taken from a simplified two-stream radiative
equilibrium model that describes how a plane-parallel atmosphere
having 
spatially constant $\gamma \equiv \kappa_{\rm V}/\kappa_{\rm IR}$, visible opacity 
$\kappa_{\rm V}$, and infrared opacity $\kappa_{\rm IR}$
absorbs incoming stellar radiation at visible wavelengths, and re-processes that energy in the thermal infrared (\citealt{pierrehumbert10}, section 4.3.5). In that model, equation (\ref{Tg}) refers specifically to the temperature of the ground, which is assumed to emit in the infrared as a perfect blackbody; any non-zero ground albedo at visible stellar wavelengths is absorbed in the parameter $c_3$. 
%assumed to emit in the thermal infrared as a blackbody (the albedo as opposed to the temperature of the atmosphere just above the ground; see \S \ref{subsec:C} where we consider the latter). 
The optical depth $\tau$ is measured from the star to the ground,
and evaluated in the infrared, i.e., in the wavebands at which the atmosphere radiates. The optical depth at visible stellar wavelengths is $\gamma \tau$.
We reiterate that
our map is not spatially resolved, so $\tau$ is not
a coordinate, but is rather
the total optical depth measured across the atmosphere (read: wind). 
%($= \tau_\infty$ in Pierrehumbert's notation).
The optical depth may take any value; equation (\ref{Tg}) is good for optically thin or thick atmospheres. Various assumptions underlying equation (\ref{Tg}) (e.g.~that the atmosphere can quickly establish radiative equilibrium) are tested in section \ref{sec:sum}.

The final component of our model relates $\tau$ to $\dot{M}$, and advances the system in time:
\begin{equation}
\tau(i+1) = c_4 \dot{M}(i) \label{tau_mdot}
\end{equation}
for $c_4 > 0$. Equation (\ref{tau_mdot}) states that the mass-loss rate from the planetary surface at time $i$ determines the optical depth of the atmosphere at a later time $i+1$ --- there is hysteresis in the system. The delay reflects the fact that the planet's wind travels at finite speed and takes time to fill whatever space it can between the planetary surface and the star. Every increment of $i$ by $+1$ advances the system by this atmospheric refresh time. From  section \ref{sec:intro}, the refresh time is of order the planet's orbital period. Thus we may also interpret each iteration of the map as taking us from one transit to the next.

Combine equations (\ref{mdot})--(\ref{tau_mdot}) into a map for $\tau$:
\begin{align}
\tau(i+1) = & \,\,c_1 c_4 \,\times\\
&\exp \{ -(c_2/c_3) [(1+1/\gamma) + (1-1/\gamma) \exp(-\gamma\tau(i))]^{-1/4} \} \,. \nonumber
\end{align}
%and $c_{4a}$ and $c_{4b}$ in equation (\ref{gamma}) for $\gamma(i) = \gamma(\tau(i))$ are the four dimensionless parameters that characterize the map.
which shows that the parameters $c_1$ and $c_4$ are degenerate, as are $c_2$ and $c_3$.
If `skies are clear' for any given time step, i.e. if $\tau(i)\ll 1$, then for the next time step $\tau(i+1) = c_1 c_4 \exp (-2^{-1/4} c_2/c_3) \equiv p_1$. We call $p_1$ the `post-clear-skies' optical depth, and  consider values over the range $p_1 \in (0.1,10)$. Further define $p_2 \equiv c_2/c_3$ to rewrite the map as:
\begin{align} \label{map}
\tau(i+1) = & \,\, p_1 \exp( +2^{-1/4} p_2 ) \, \times \\
& \exp \{ -p_2 [(1+1/\gamma) + (1-1/\gamma) \exp(-\gamma\tau(i))]^{-1/4} \} \,. \nonumber
\end{align}
We decide the range for $p_2$ 
as follows. From \cite{perezbecker13}, $c_2$ in the Clausius-Clapeyron equation (\ref{mdot}) varies from $6.9777 \times 10^4$ K (pyroxene) to $6.5649 \times 10^4$ K (olivine) to $4.2694 \times 10^4$ K (iron).
\cite{perezbecker13} further estimate the peak dayside temperature of KIC 1255b under optically thin conditions to be $T_0 = 2150$ K (their equation 6); this implies $c_3 = 2^{-1/4} T_0 = 1808$ K from equation (\ref{Tg}). From these considerations, $p_2 = c_2/c_3 \in (23.615, 38.595)$. We widen the range surveyed to $p_2 \in (20,40)$.

In section \ref{subsec:A}, we explore the map (\ref{map}) assuming the opacity ratio $\gamma$ is constant in time. These results motivate
a model where $\gamma$ varies with $\tau$, presented in section \ref{subsec:B}. We construct a third map in section \ref{subsec:C} by
replacing equation (\ref{Tg})
with an alternative prescription
for the surface temperature.

\subsection{Map A: Constant $\gamma$} \label{subsec:A}

\begin{figure}
\includegraphics[width=1.0\linewidth]{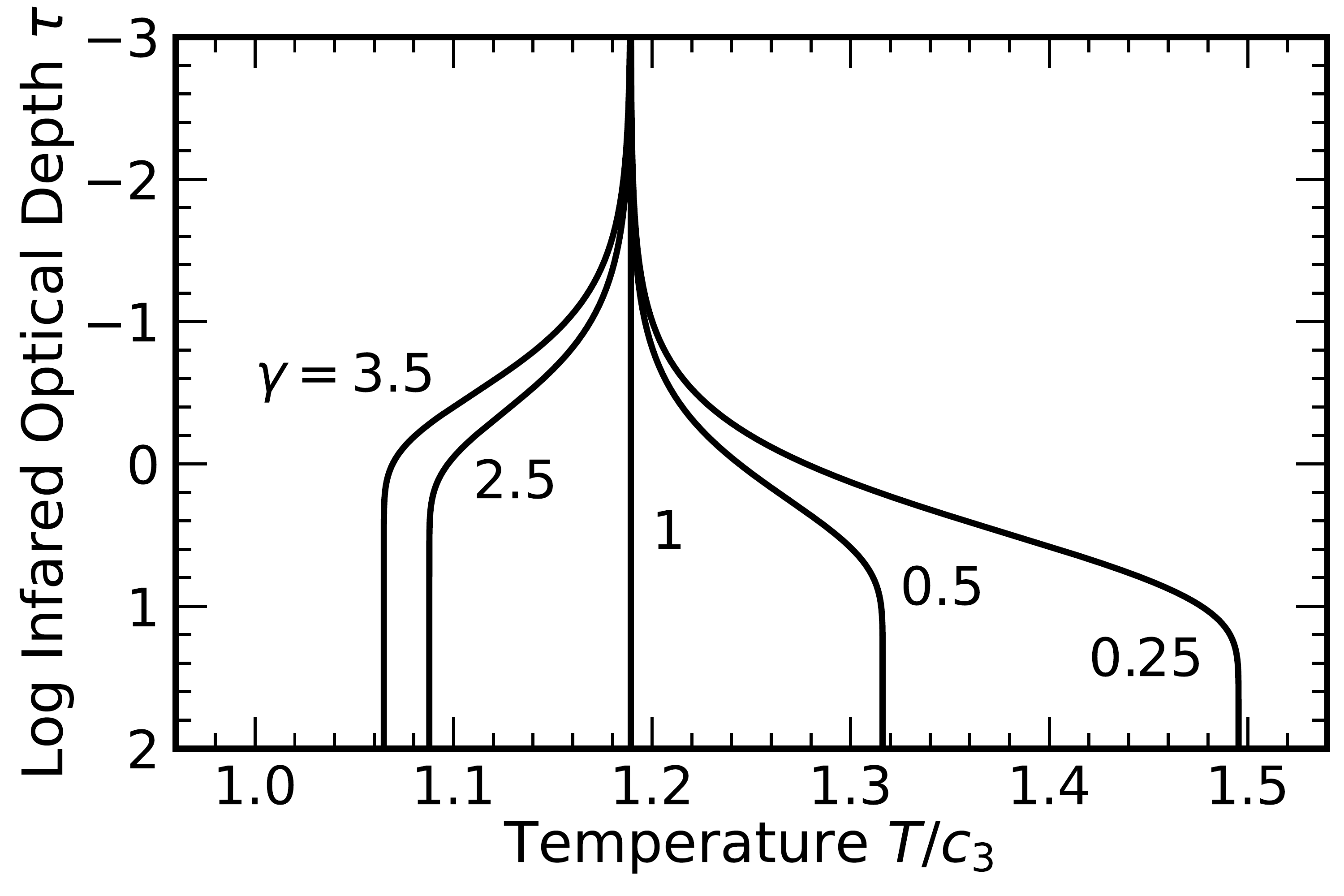}
\caption{Ground temperature $T$ vs. infrared optical depth $\tau$, from equation (\ref{Tg}) for an atmosphere in radiative equilibrium. Depending on the visible-to-infrared opacity ratio $\gamma$, the ground either increases in temperature as the atmosphere becomes optically thicker ($\gamma < 1$ `greenhouse'), or decreases in temperature ($\gamma > 1$ `nuclear winter'). Each curve in this figure is computed assuming $\gamma$ = constant; we keep this assumption in Map A but relax it for Maps B and C.}
%Fig 1
\label{fig:Ttau}
\end{figure}

Equation (\ref{Tg}) derives from a model of an atmosphere in radiative
equilibrium that behaves qualitatively differently depending on
whether the optical-to-infrared opacity ratio $\gamma$
is less than or greater than 1 --- see Figure \ref{fig:Ttau}. 
When $\gamma < 1$, starlight at visible
wavelengths penetrates the atmosphere more easily than re-processed infrared radiation can escape. The result is a `greenhouse effect' whereby the temperature increases toward the planet's surface. The surface temperature $T$ increases with increasing total atmospheric optical depth $\tau$, up to a theoretical maximum that depends on $\gamma$ (Fig.~\ref{fig:Ttau}).

These trends reverse when $\gamma > 1$. In this regime the atmosphere has an `inversion layer'; most of the incident visible radiation is absorbed by and
heats the uppermost layers of the atmosphere, which directs just a
portion of the re-processed infrared radiation downward to heat the planetary
surface. The more opaque the atmosphere, the less the planet's surface
is heated --- but just as there is a ceiling on $T$ in the $\gamma < 1$ greenhouse, there is a floor on $T$ in this $\gamma > 1$ `nuclear winter' (Fig.~\ref{fig:Ttau}; \citealt{pierrehumbert10}).

\begin{figure}
\centering 
\includegraphics[width=1.0\linewidth]{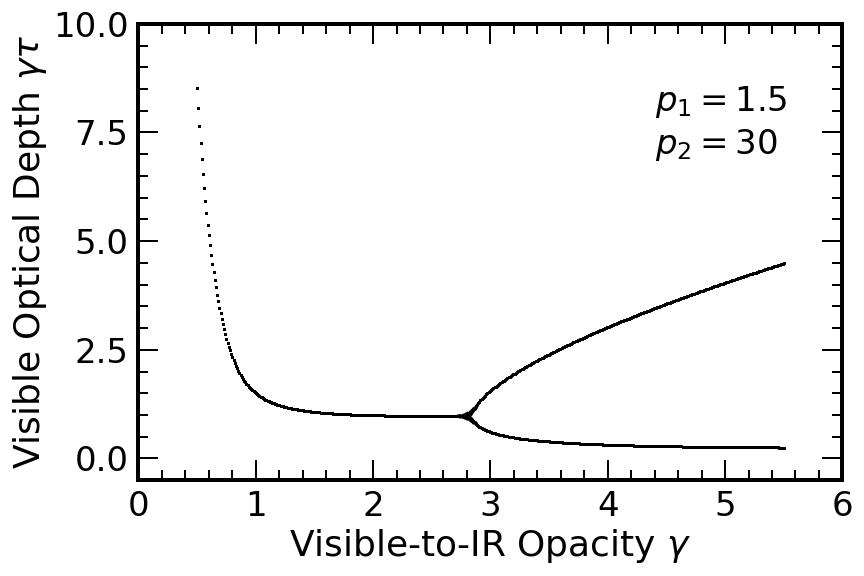}
\caption{An orbit diagram (e.g. \citealt{strogatz15}) for Map A (equation \ref{map} for fixed $\gamma$), assuming $p_1 = 1.5$ and $p_2 = 30$. For a given opacity ratio $\gamma$, we iterate Map A 300 times starting from $\tau(i=0) = 0$, and plot the values of $\tau(i)$ (multiplied by $\gamma$ to give the visible optical depth rather than the infrared optical depth) after discarding the first 100 iterations. For $\gamma < 2.8$, the map converges to a single-valued equilibrium. For $\gamma > 2.8$, the map bifurcates into a 2-cycle that alternates between optically thin and optically thick states.}
%Fig 2
\label{fig:orbitDiagram}
\end{figure}

Figure \ref{fig:orbitDiagram} is an orbit diagram (\citealt{strogatz15}) showing how the results of the map (\ref{map}) for $\tau$ vary with the parameter $\gamma$, at fixed $p_1 = 1.5$ and $p_2 = 30$. We plot the visible optical depth $\gamma\tau$ which is more relevant for the {\it Kepler} spacecraft data. For a given small $\gamma$,
the map converges to a single fixed point, $\gamma\tau = \gamma \tau_{\rm f}(\gamma)$. By contrast, for large $\gamma \gtrsim 2.8$, the solution bifurcates into a
2-cycle that alternates between small $\gamma\tau$ and moderate $\gamma\tau$. 

\begin{figure*}
    \centering
    \includegraphics[width = 1.0 \textwidth]{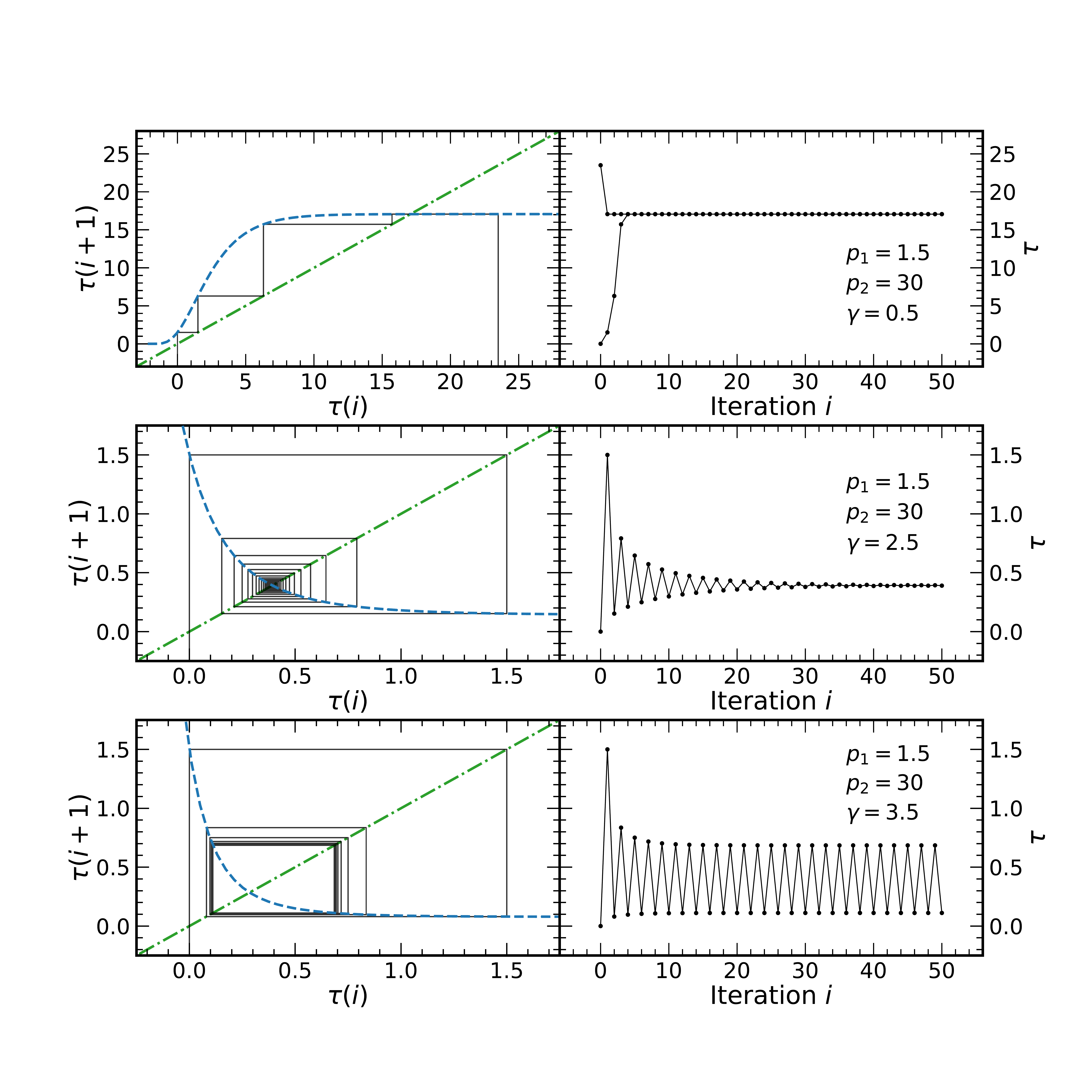}
    \caption{{\it Right:} Sample time series for Map A for three different values of $\gamma$ (top to bottom). The map converges to a single-valued equilibrium for $\gamma = 0.5$ and 2.5, and a 2-cycle for $\gamma = 3.5$ (see also~Fig.~\ref{fig:orbitDiagram}). We initialize our maps with $\tau(i=0) = 0$ except in the top panels  where we experiment with $\tau(i=0) = 23.5$ to demonstrate convergence to equilibrium from above. {\it Left:} Cobweb diagrams (\citealt{strogatz15}) illustrating system trajectories. The map from $\tau(i)$ to $\tau(i+1)$ --- what we also call the ``iteration curve'' --- is plotted as a blue dashed curve, while the ``1-1'' line $\tau(i) = \tau(i+1)$ is plotted as a green dot-dashed curve. Each trajectory is plotted as a black solid curve which alternates between traveling vertically from the 1-1 line to the iteration curve, to traveling horizontally from the iteration curve to the 1-1 line. Intersections of the iteration curve with the 1-1 line are fixed points (a system that starts there stays there); the fixed points are stable against small perturbations if the 
 local slope of the iteration curve is $> -1$ and $<1$, and unstable otherwise. For $\gamma = 0.5$ and $2.5$, the fixed points are stable and act as attractors. For $\gamma = 3.5$, the fixed point is unstable and the trajectory encircles it in a 2-cycle.}
 %Fig 3
    \label{fig:tsGamma}
\end{figure*}

The single-valued equilibrium at low $\gamma$ and its bifurcation at high $\gamma$ can be
understood by examining the shape of the $\tau(i+1)$ vs. $\tau(i)$
curve --- hereafter the `iteration curve', synonymous with the map --- and how it varies with
$\gamma$. 
We sample three iteration curves for $\gamma = \{0.8, 2.5, 3.5\}$ in Figure \ref{fig:tsGamma}, for the same values of $p_1$ and $p_2$ as in Fig.~\ref{fig:orbitDiagram}. 
Wherever the iteration curve (blue dashed) intersects the $\tau(i+1) =
\tau(i)$ line (green dot-dashed),
there is an equilibrium, a.k.a. a fixed point. For Map A which assumes constant $\gamma$,
the iteration curve is shaped such that there is only
one fixed point, at a value $\tau = \tau_{\rm f}$ that depends on $\gamma$ and the other parameters.

A fixed point is linearly stable
if the slope of the iteration curve at that point lies between $-1$ and $1$;
%$> -1$; 
otherwise it is unstable (\citealt{strogatz15}). The bifurcation at $\gamma \simeq 2.8$
in Fig.~\ref{fig:orbitDiagram} divides stable points at $\gamma < 2.8$ from
unstable points at $\gamma > 2.8$. When $\gamma < 1$ (greenhouse), 
the iteration curve monotonically increases; its slope is
always positive but never exceeds $+1$ at the fixed point $\tau_{\rm f}$, which is therefore
%, and so the fixed point $\tau_{\rm f}$ is 
always stable (top row of Fig.~\ref{fig:tsGamma}).
The map converges to this stable fixed point starting from either side of
$\tau_{\rm f}$. We can understand this behaviour as follows. If $\tau < \tau_{\rm f}$, the atmosphere becomes
progressively dustier and warmer from the greenhouse effect until it reaches $\tau_{\rm f}$. If
$\tau > \tau_{\rm f}$, the surface temperature $T$ is nearly saturated at its theoretical ($\gamma$-dependent) 
maximum (Fig.~\ref{fig:Ttau}), which means the mass-loss and
dust-production rates have also saturated, at values that cannot
sustain the given $\tau$. Thus on the next iteration $\tau$ drops, back toward $\tau_{\rm f}$.

If $\gamma > 1$ (nuclear winter), the iteration curve monotonically decreases (middle and bottom rows of Fig.~\ref{fig:tsGamma}). But only for $\gamma$ sufficiently large does
the iteration curve slope at $\tau_{\rm f}$ become sufficiently negative to destabilize the equilibrium there (bottom row). The resultant 2-cycle flips back and forth across the unstable fixed point. When $\tau$ is small, $T$ is relatively high (Fig.~\ref{fig:Ttau}), as is $\dot{M}$ by extension; this leads to higher $\tau$ in the next timestep, and thus lower $T$ and lower $\dot{M}$, re-starting the cycle. We see that 2-cycles require not only that $T$ decrease with increasing $\tau$ ($\gamma > 1$), but that the decrease of $T$ with $\tau$ be sufficiently large to transform the fixed point from attractive (middle panel) to repulsive (bottom panel).

Bifurcation to a 2-cycle does not occur for all $p_1$ and $p_2$.
The slope of the iteration curve, which needs to be sufficiently negative to destabilize the fixed point, is the derivative of equation (\ref{map}) with respect to $\tau(i)$. It is clear, for example, that too small a value of $p_1$ prevents destabilization. Our numerical experiments with Map A over parameter space yield either single-valued attractors for all $\gamma$, or single-valued attractors at small $\gamma$ and a 2-cycle at large $\gamma$.

%EC: A relevant post: https://physics.stackexchange.com/questions/331195/if-f-cdot-is-a-chaotic-map-then-is-f-1-cdot-known-as-the-inverse-map
Map A does not generate chaotic trajectories. This is to be expected, since the $T(\tau)$ relation for constant $\gamma$ is monotonic (Fig.~\ref{fig:Ttau}), leading to an iteration curve which is also monotonic and therefore invertible (Fig.~\ref{fig:tsGamma}). Invertible one-dimensional maps are regular. In the next section \ref{subsec:B}, we relax the assumption of constant $\gamma$ to see if chaos might result.

\subsection{Map B: $\gamma(\tau)$} \label{subsec:B}

\subsubsection{Rationale and construction}\label{subsubsec:B}
Map A assumes that the optical-to-infrared opacity ratio $\gamma$ is fixed for all time, i.e. the optical properties of the dust grains condensing out of the planet's wind and dominating the opacity are assumed invariant. Here for Map B we relax this assumption, considering how young grains newly condensed under clear skies may differ systematically from more evolved grains growing under dustier conditions.

Motivated by ideas in \citet[][their section 3.2]{booth23}, we suppose $\gamma < 1$ when the visible optical depth $\gamma\tau \ll 1$. Under optically thin (clear sky) conditions, the full stellar irradiance, concentrated at visible wavelengths, prevents condensates from forming if they contain minerals that absorb too strongly in the visible and emit too weakly in the infrared; such grains would be heated to temperatures too high to remain in solid form. Thus we expect grains that condense under optically thin conditions to have $\gamma < 1$. \citet{booth23} propose that freshly condensed grains would be iron-poor and silicate-rich, absorbing poorly in the visible and more strongly in the infrared (e.g. at a wavelength of $\sim$10 $\mu$m from the silicate waveband). This expectation holds for practically all grain sizes, from $10^{-2}$ $\mu$m on up (see figure 4 of \citealt{booth23}).

Conversely, under optically thick conditions ($\gamma\tau \gg 1$), grains would be shielded from direct visible radiation, and may have $\gamma \gtrsim 1$. \citet{booth23} suggest that as grains grow in size, they can accrete iron by heterogeneous nucleation, and thereby absorb more strongly in the optical. Most grains in the universe have $\gamma > 1$ (e.g. \citealt{draine11}).

\begin{figure}
\centering 
%\vspace{-0.75in}
\includegraphics[width=1.0\linewidth]{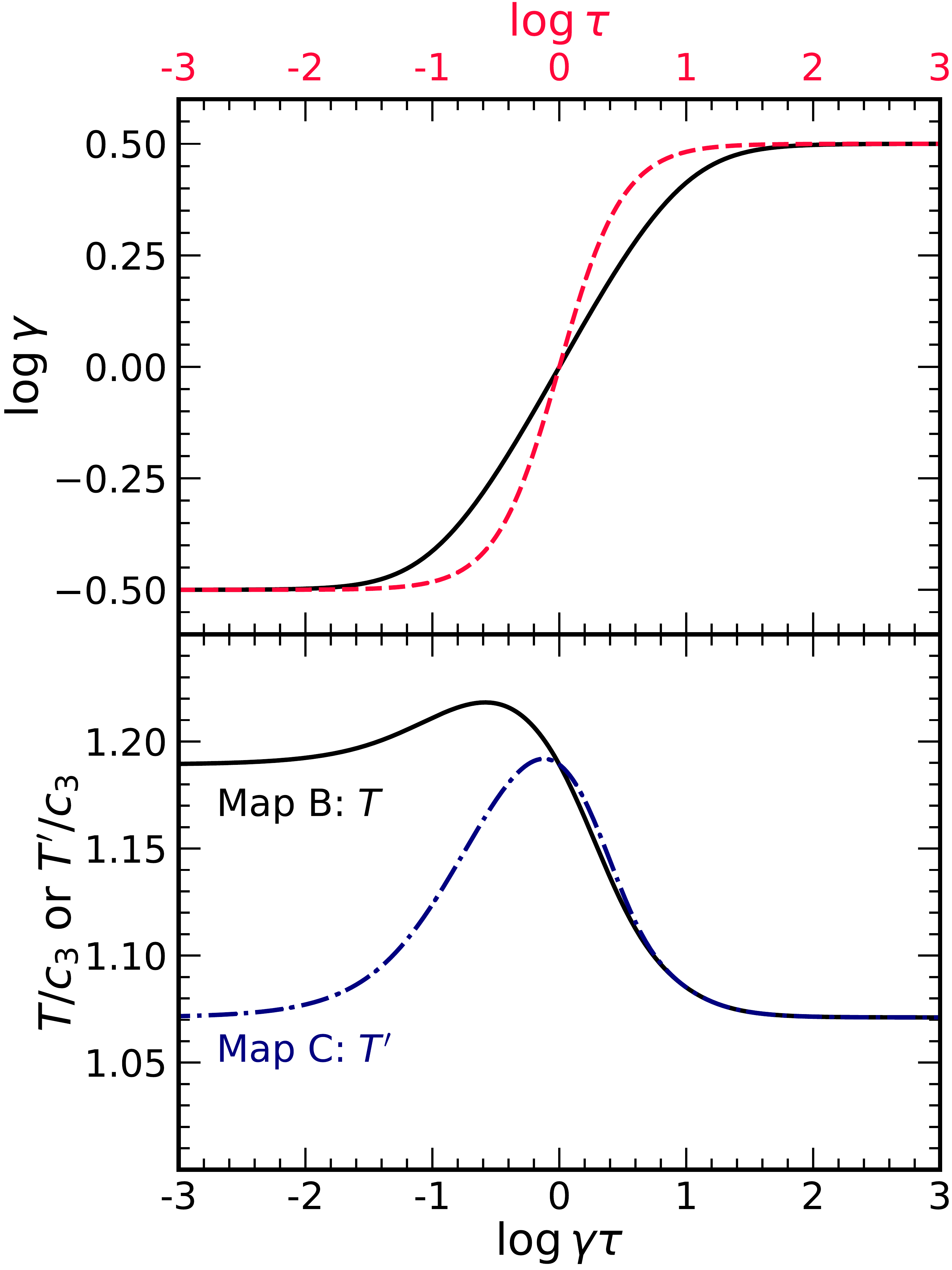}
\vspace{-0.0in}
\caption{{\it Top:} Visible-to-infrared opacity ratio $\gamma$ vs. 
infrared optical depth $\tau$ (red, top axis) and
visible optical depth $\gamma\tau$ (black, bottom axis) according to equation (\ref{gamma}) with $p_3 = 0.5$ and $p_4 = 0.5$. {\it Bottom:} Normalized ground temperature vs. visible optical depth $\gamma(\tau)\tau$ for Maps B (black, $T/c_3$) and C (blue, $T'/c_3$) according to equations (\ref{Tg}) and (\ref{Tg2}), respectively. Allowing $\gamma$ to vary with $\tau$, as in the top panel, leads to temperature varying non-monotonically with $\tau$ (cf. Fig. \ref{fig:Ttau}).}
%Fig 4
\label{fig:Tg}
\end{figure}

We conscript the $\tanh$ function to switch between the two regimes, prescribing $\gamma$ to vary with $\tau$ according to
\begin{align} \label{gamma}
\log_{10}\gamma(\tau) = p_{3} \tanh [(\log_{10} \tau)/p_{4}]
\end{align}
for free parameters $p_3, p_4 > 0$. When $\tau \ll 1$, $\gamma \rightarrow 10^{-p_{3}}$, and when $\tau \gg 1$, $\gamma \rightarrow 10^{+p_{3}}$. The parameter $p_{4}$ controls how rapidly this switch is made in $\log_{10}\tau$ space. As seen in Figure \ref{fig:Tg}, the switch occurs more gradually in visible-wavelength optical depth $\gamma \tau$ than in infrared optical depth $\tau$. The visible-wavelength optical depth is more physically relevant than the infrared insofar as the former underlies our narrative about grain condensation. Nonetheless we will often plot the infrared $\tau$ for simplicity (equation \ref{map} is a map for $\tau$), knowing that $\tau$ can always be converted to $\gamma\tau$ via equation (\ref{gamma}). We consider $p_3 \in (0.1,2)$ and for simplicity fix $p_4 = 0.5$.

Having $\gamma$ vary with $\tau$ in this way leads to a non-monotonic $T(\tau)$ relation (Fig.~\ref{fig:Tg}, bottom panel): $T$ grows with $\gamma\tau$ when $\gamma\tau \lesssim 1$ (greenhouse heating), and then decreases with $\gamma\tau$ when $\gamma\tau \gtrsim 1$ (inversion layer cooling). Contrast this behaviour with the monotonic $T(\tau)$ relations for constant $\gamma$ in Fig.~\ref{fig:Ttau}. A non-monotonic $T(\tau)$ curve can lead to a non-invertible map for $\tau$, and a non-invertible one-dimensional map can exhibit chaos.

\subsubsection{Results for Map B}\label{subsubsec:results_B}
\begin{figure}
%\centering 
%%\vspace{-0.75in}
\includegraphics[width=1.0\linewidth]{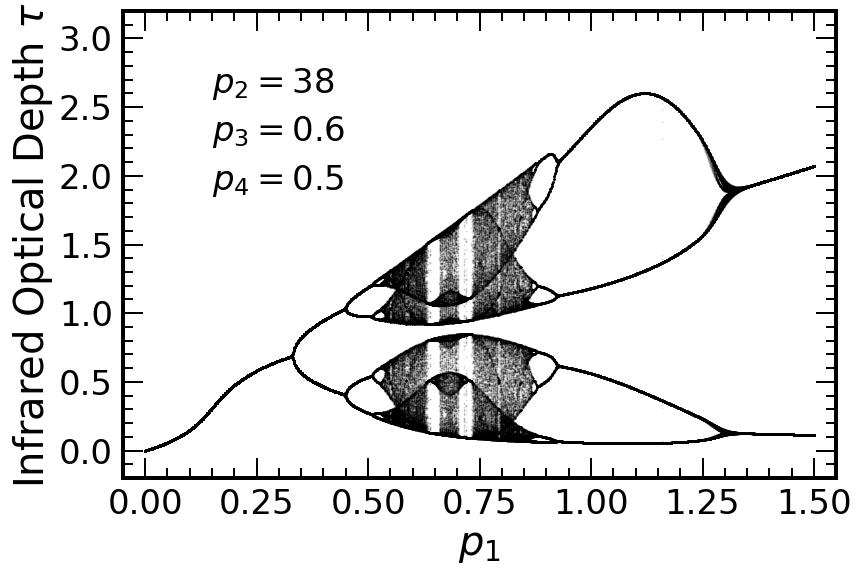}
\vspace{-0.0in}
\caption{An orbit diagram (see Fig. \ref{fig:orbitDiagram}) for Map B assuming $p_2 = 38$, $p_3 = 0.6$, and $p_4 = 0.5$. For a given $p_1$ we iterate Map B 200 times starting from each of 5 values evenly spaced from $0 \leq \tau(i=0) \leq 3$ and plot the last 100 values of $\tau(i)$. 
%We repeat this with with diffierent initial value for a total of 5 evenly spaced initial values in the range $0 \leq \tau(i = 0) \leq 3$. 
The map bifurcates repeatedly from $p_1=0$ to $p_1 \approx 0.5$ until it becomes chaotic for $p_1 \gtrsim 0.5$. The chaos is interrupted by two ``periodic windows'' (\citealt{strogatz15}) centered around $p_1 \approx 0.65$ and $p_1 \approx 0.72$.  For $p_1 \gtrsim 0.9$, the map reverts back to regular limit-cycle behaviour.}
%Fig 5
\label{fig:orbitDiagramA}
\end{figure}

\begin{figure}
\centering 
\includegraphics[width=1.0\linewidth]{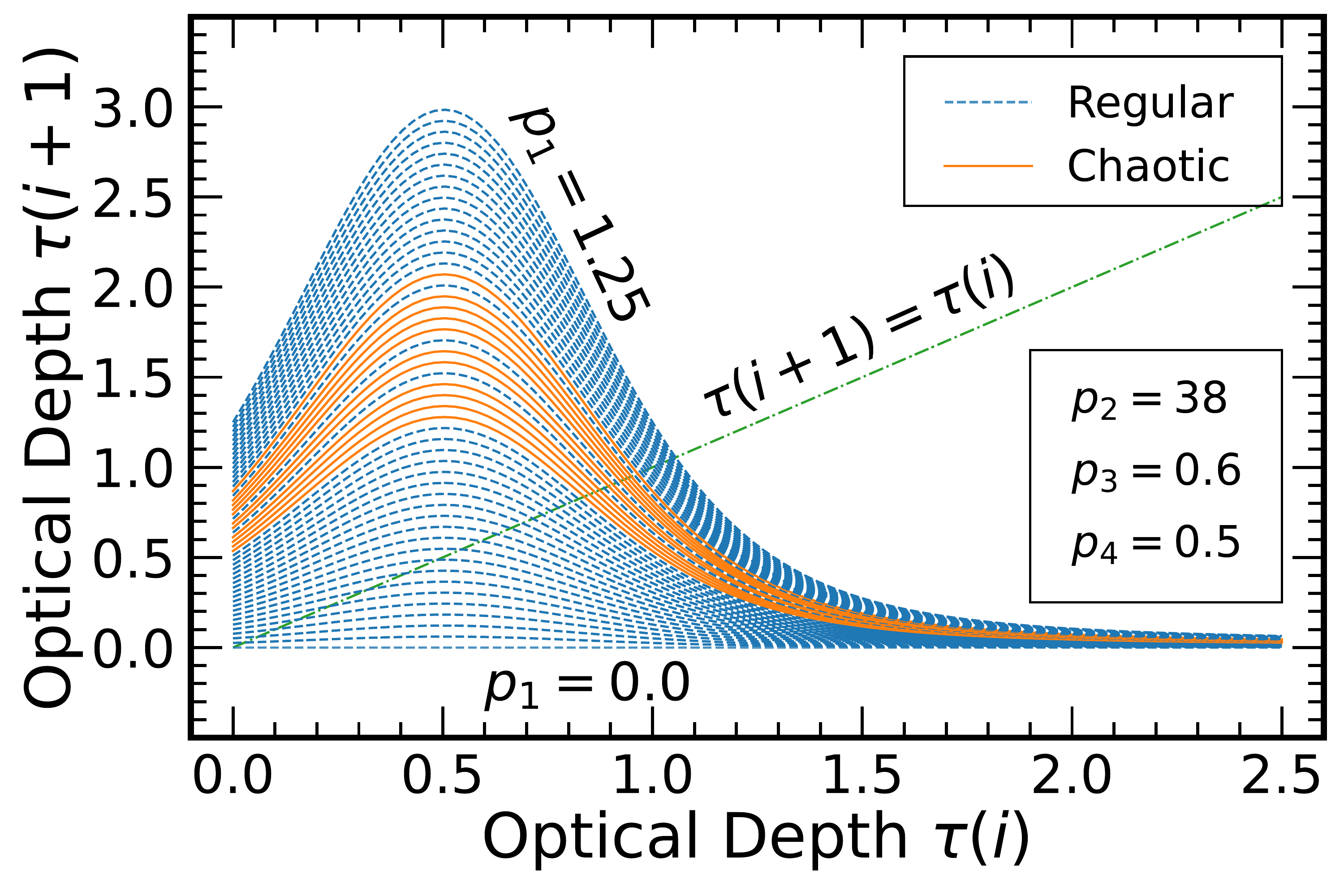}
\vspace{-0.0in}
\caption{A family of Map B iteration curves for $p_2 = 38$, $p_3 = 0.6$ and $p_4 = 0.5$ with $p_1$ increasing from 0 to 1.5 in steps of $0.03$. Blue iteration curves yield regular behaviour, either periodic limit cycles or single-valued equilibria (1-cycles). Orange curves produce chaos. Compare with Fig.~\ref{fig:orbitDiagramA}.} 
%Fig 6
\label{fig:mapFamilyA}
\end{figure}

The potential for Map B to produce chaos is realized in Figure \ref{fig:orbitDiagramA} showing an orbit diagram parameterized by $p_1$ (the post-clear-sky optical depth); other parameters are held fixed for now at $p_2 = 38$ and $p_3 = 0.6$. 
The underlying family of iteration curves (maps) is shown in Figure \ref{fig:mapFamilyA}, color coded by whether they produce regular (blue dashed) or chaotic (orange solid) trajectories. The iteration curves are all non-monotonic, reflecting the non-monotonicity of $T(\tau)$ (Fig.~\ref{fig:Tg}). Each map rises from greenhouse warming at low $\tau$, and falls from nuclear winter cooling at high $\tau$, the amplitude of variation increasing with $p_1$. 

\begin{figure*}
\centering 
%\vspace{-0.75in}
\includegraphics[width=0.8\textwidth]{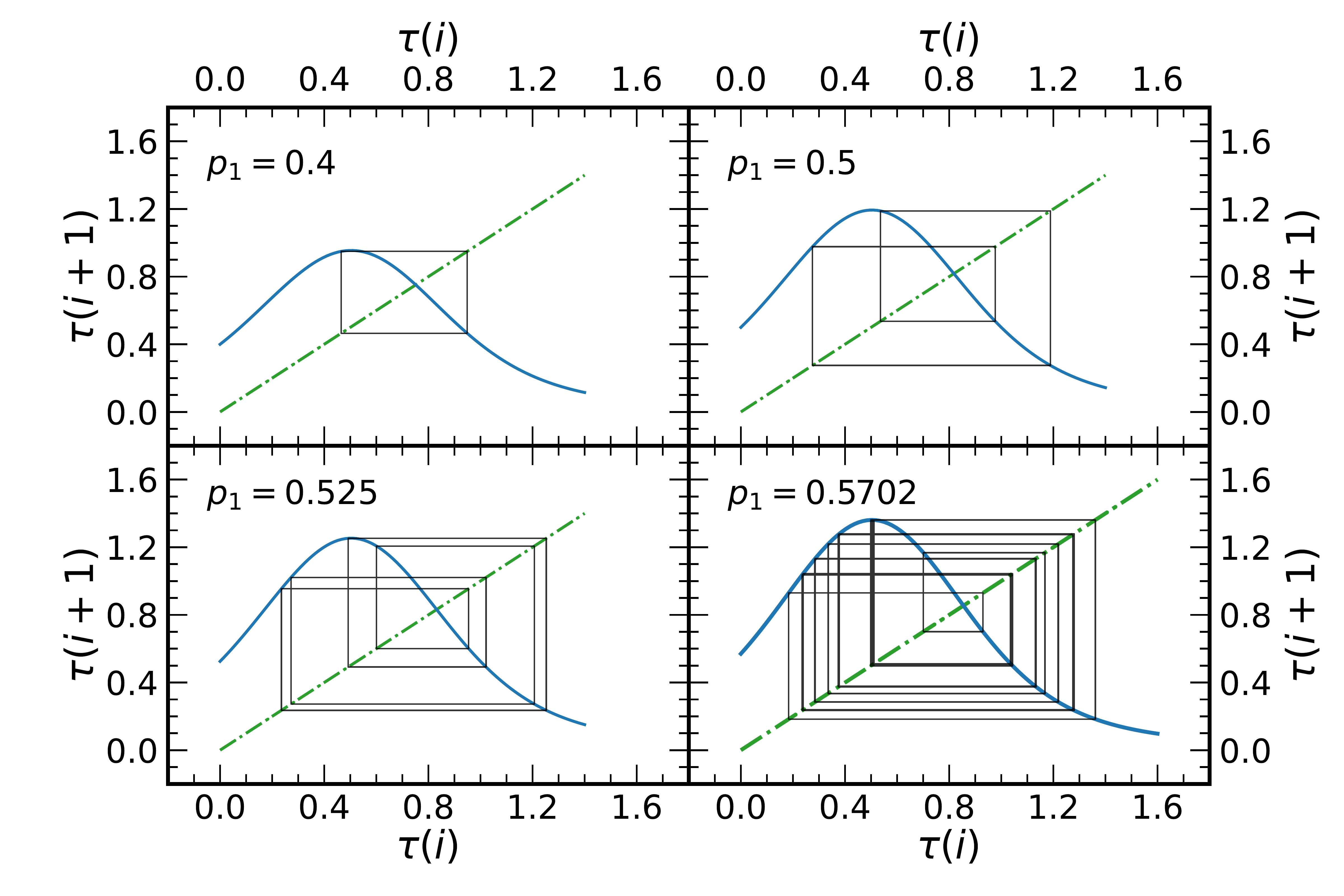}
\vspace{-0.0in}
\caption{Cobwebs for Map B with $p_2 = 38$, $p_3 = 0.6$, and $p_4 = 0.5$, illustrating repeated bifurcations with increasing $p_1$. Iteration curves are plotted in blue, 1:1 lines in green, and trajectories in black, obtained by iterating the map 200 times starting from $\tau(0) = 0$ and discarding the first 100 iterations (see caption to Fig.~\ref{fig:tsGamma} to see how cobwebs are constructed). {\it Top left:} A 2-cycle for $p_1$ = 0.4. {\it Top right:} A 4-cycle for $p_1 = 0.5$. {\it Bottom left:} An 8-cycle for $p_1 = 0.525$. {\it Bottom right:} A cycle with around 20 points for $p_1 = 0.57$.}
%Fig 7
\label{fig:cobwebs}
\end{figure*}

The rise and fall of Map B in Fig.~\ref{fig:mapFamilyA} recalls the shape of the quadratic logistic map $f(x) = rx(1-x)$ (\citealt{may76}; \citealt{strogatz15}), with $p_1$ playing the role of the growth rate $r$. Like the logistic map, Map B yields an orbit diagram that bifurcates repeatedly until dissolving into a sea of chaos (Fig.~\ref{fig:orbitDiagramA}). Each bifurcation is triggered by a fixed point of the $n^{\rm th}$-iterate map, either
$f^n(x)$ for the logistic map or $\tau^n(i)$ for Map B, becoming unstable. We can see the first-iterate fixed point of Map B becoming unstable in Fig.~\ref{fig:mapFamilyA}: as $p_1$ increases, the slope of the iteration curve at the fixed point becomes increasingly negative, falling below -1 at $p_1 \simeq 0.33$, just below the first orange curve from the bottom --- this marks the first bifurcation in Fig.~\ref{fig:orbitDiagramA}. The resultant 2-cycle around the unstable fixed point alternates between a greenhouse and a nuclear winter. Each bifurcation doubles the number
of points in a cycle and the cycle period (Figure~\ref{fig:cobwebs}), with successive bifurcations unfurling increasingly rapidly with $p_1$.%\footnote{From Fig.~5, the intervals $\Delta p_1$ between the first and second bifurcations, the second and third bifurcations, and third and fourth bifurcations are 0.12, 0.06, and 0.02, respectively. From these we compute the first two terms in the Feigenbaum sequence (e.g. Strogatz 2000) for Map B, 0.12/0.06 = 2 and 0.06/0.02 = 3. The analogous sequence for the logistic map converges to approximately 4.669, a.k.a. Feigenbaum's constant. We do not expect the sequence for Map B to converge to this same classic value, since the derivation of Feigenbaum's constant relies on the map intersecting the origin, which Map B does not.}

\begin{figure*}
\centering 
%\vspace{-0.75in}
\includegraphics[width=0.8\textwidth]{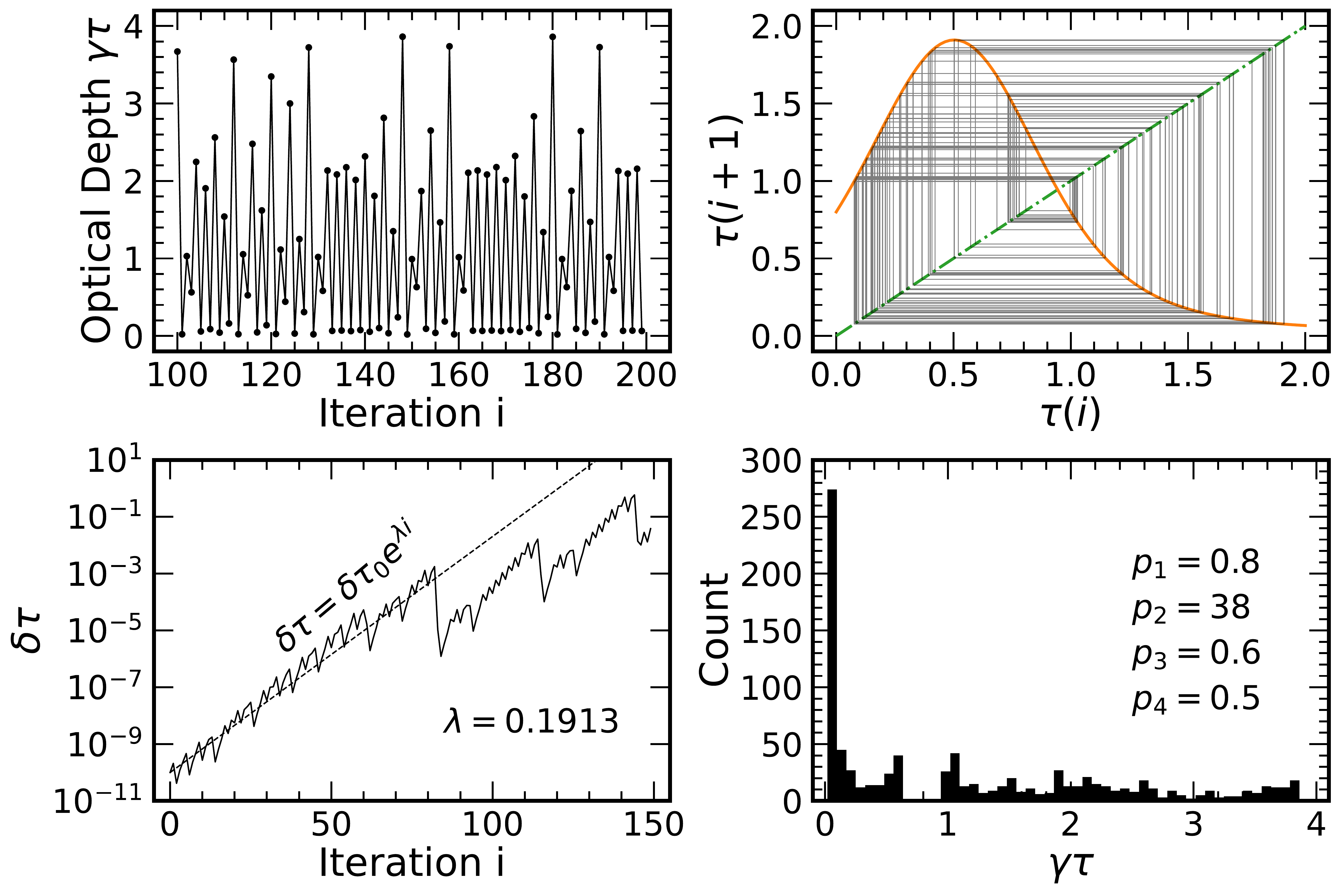}
\vspace{-0.0in}
\caption{
{\it Top left:} Visible-wavelength optical depth for Map B with $p_1 = 0.8$, $p_2 = 38$, $p_3 = 0.6$ and $p_4 = 0.5$. The map was initialized with $\tau(i=0)=0$ and the first 100 iterations discarded. {\it Top right:} Same trajectory as in top left, plotted in cobweb form. The iteration curve is plotted in orange to signify that it produces chaos (same color coding as in Fig.~\ref{fig:mapFamilyA}). {\it Bottom left:} Neighboring trajectories diverge exponentially with a Lyapunov exponent of $\lambda = 0.1913$ iteration$^{-1}$. The variable $\delta \tau$ is the difference in $\tau$ values between two trajectories with an initial difference of $\delta \tau_0 = 10^{-10}$. The difference $\delta \tau$ eventually stops following an exponential curve (dashed line) since trajectories are bounded (here to $\tau < 2$). 
{\it Bottom right:} A histogram of all points visited by the map over 1000 iterations with the first 100 iterations discarded.
%The histogram is divided into a total of 50 bins.
}
%Fig 8
\label{fig:timeSeries8}
\end{figure*}

\begin{figure*}
\centering 
%\vspace{-0.75in}
\includegraphics[width=0.8\textwidth]{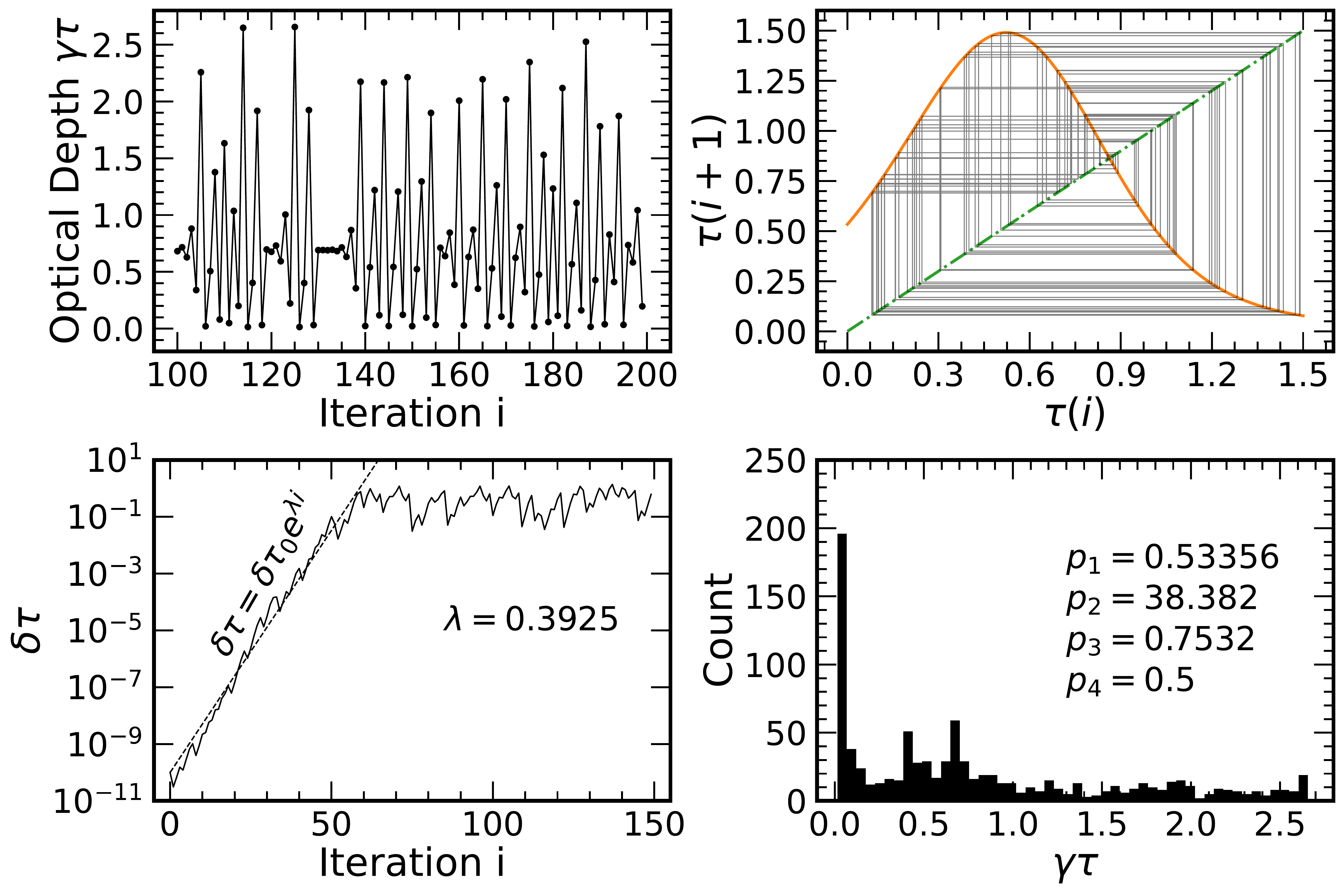}
\vspace{-0.0in}
\caption{Same as Fig.~\ref{fig:timeSeries8} but for a more chaotic trajectory (higher Lyapunov exponent $\lambda = 0.3925$ iteration$^{-1}$) that does not avoid the unstable fixed point as much. Map parameters are $p_1 = 0.53356$, $p_2 = 38.382$, $p_3 = 0.7532$, and $p_4 = 0.5$.}
%Fig 9
\label{fig:timeSeries94}
\end{figure*}

Once $p_1 \gtrsim 0.5$, the trajectory has bifurcated so many times that it wanders through a quasi-continuum of values: Map B has become chaotic. We can see how the chaos plays out in Figure~\ref{fig:timeSeries8} which shows a sample trajectory having a Lyapunov exponent of $0.19$ per iteration; a trajectory that starts infinitesimally close to the one shown diverges exponentially from it with an e-folding time of $0.19^{-1} \simeq 5$ iterations. 
This particular trajectory avoids the unstable fixed point at $\tau \simeq 0.9$. The same avoidance can be seen in the orbit diagram of Fig.~\ref{fig:orbitDiagramA}, showing the chaotic sea parted by the locus of fixed points (the locus traced by the green dot-dashed line in Fig.~\ref{fig:mapFamilyA} and its map intersections). 
Such avoidance is not universally seen. Figure \ref{fig:timeSeries94} shows a trajectory in a different portion of $\{p_1,p_2,p_3\}$ space; it lands frequently near the fixed point, so close that it spends considerable time spiraling away. The trajectory in Fig.~\ref{fig:timeSeries94} is more chaotic than the one in Fig.~\ref{fig:timeSeries8}.

For the highest values of $p_1$, 
%large $p_1 \gtrsim 0.85$, 
the system returns to a regular 2-cycle (Figs.~\ref{fig:orbitDiagramA} and \ref{fig:mapFamilyA}): a clear sky is followed immediately by a wind so strong and dusty (by virtue of large $p_1$) that the planet plunges into a cold nuclear winter, which shuts off the wind and re-starts the cycle.

\begin{figure*}
\centering 
%\vspace{-0.75in}
\includegraphics[width=1.0\linewidth]{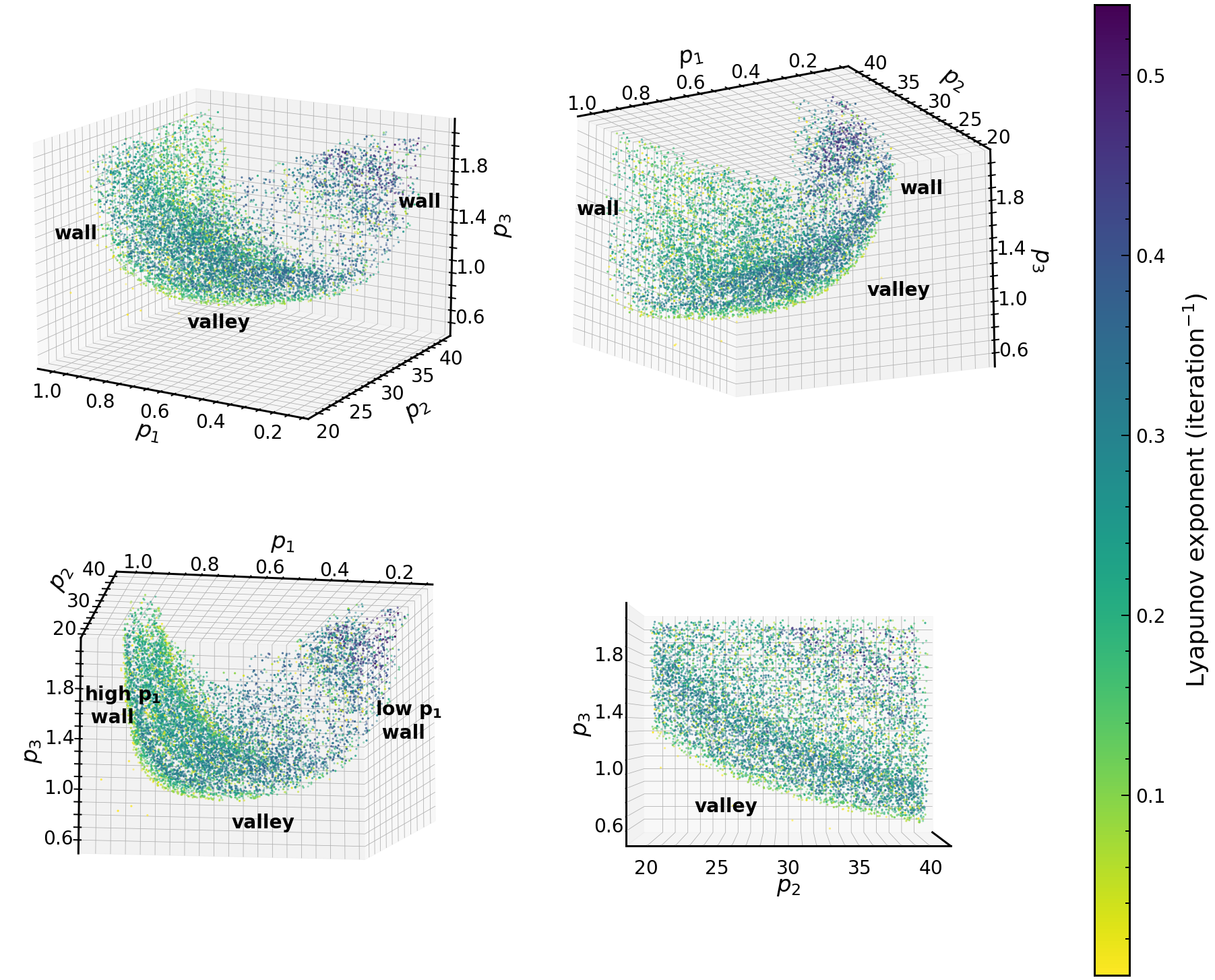}
\vspace{-0.0in}
\caption{Chaotic locus of Map B. At each of 50000 points randomly
  distributed in the 3D space of $0 < p_1 < 1$, $20 < p_2 < 40$ and $0 < p_3 <
  2$ ($p_4 =0.5$ is fixed), the Lyapunov exponent is calculated; if positive, that point is
  marked on the plot, colored according to the magnitude of the
  exponent. All panels show the same data viewed from different angles around the $p_3$ axis. The two `walls' and the `valley' in between are features discussed in the main text.}
%A plot of the chaotic locus of map B. 50000 points were randomly generated within $0 < p_1 < 1$, $20 < p_2 < 40$ and
         %$0 < p_3 < 2$ with $p_4 = 0.5$. A lyapunov exponent was
         %calculated for each of these points and all point with a
         %positive lyapunov exponent were plotted. The points are
         %colored according to their lyapunov exponent. The four
         %panels show view of the chaotic locus from different
         %angles.
%Fig 10
\label{fig:locus1}
\end{figure*}

\begin{figure}
\centering
\includegraphics[width=1.0\linewidth]{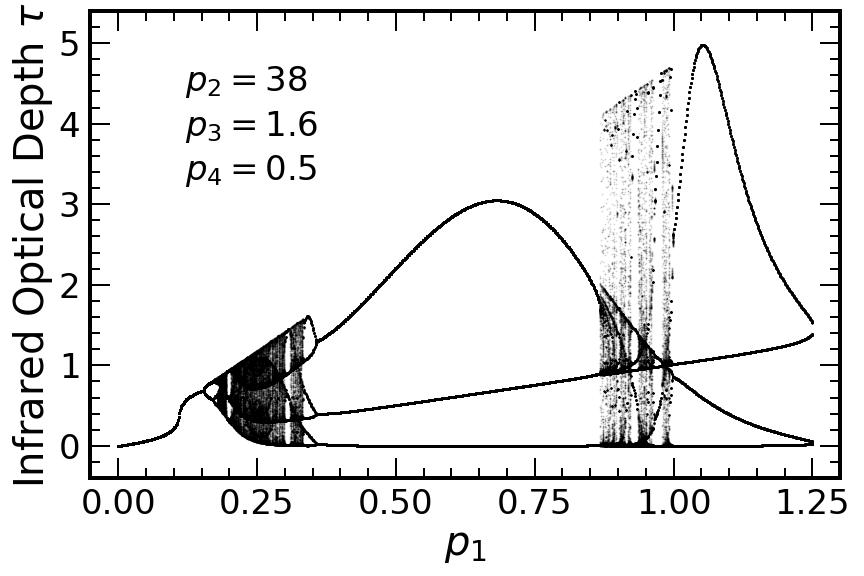}
\vspace{-0.0in}
\caption{Orbit diagram for Map B at relatively high $p_3 =
  1.6$, with $p_2 = 38$ and $p_4 = 0.5$. The first chaotic region at low
  $p_1 \sim 0.2$ is marked by successive bifurcations and appears akin to
  the chaotic region plotted in Fig.~\ref{fig:orbitDiagramA}. The
  second chaotic region at high $p_1
  \sim 0.9$ appears qualitatively distinct. Here $p_1$ (the
  post-clear-skies optical depth) nearly
  coincides with the unstable fixed point; the system 
  repeatedly lands near and spirals away from this point (see Figure \ref{fig:timeSeries9} for an illustration). To construct this plot, we iterate Map B 200 times for a given $p_1$ starting from each of 5 values of $\tau(i=0)$ spaced evenly from 0 to 5 inclusive and plot the last 100 values of $\tau(i)$. 
%We repeat this with different initial values for a total of 5 evenly spaced initial values in the range $0 \leq \tau(i=0) \leq 5$
%Orbit Diagram for $p_2 = 38$, $p_3 = 1.6$, $p_4 = 0.5$.
}
%Fig 11
\label{fig:orbitDiagramB}
\end{figure}

\begin{figure}
\centering 
%\vspace{-0.75in}
\includegraphics[width=1.0\linewidth]{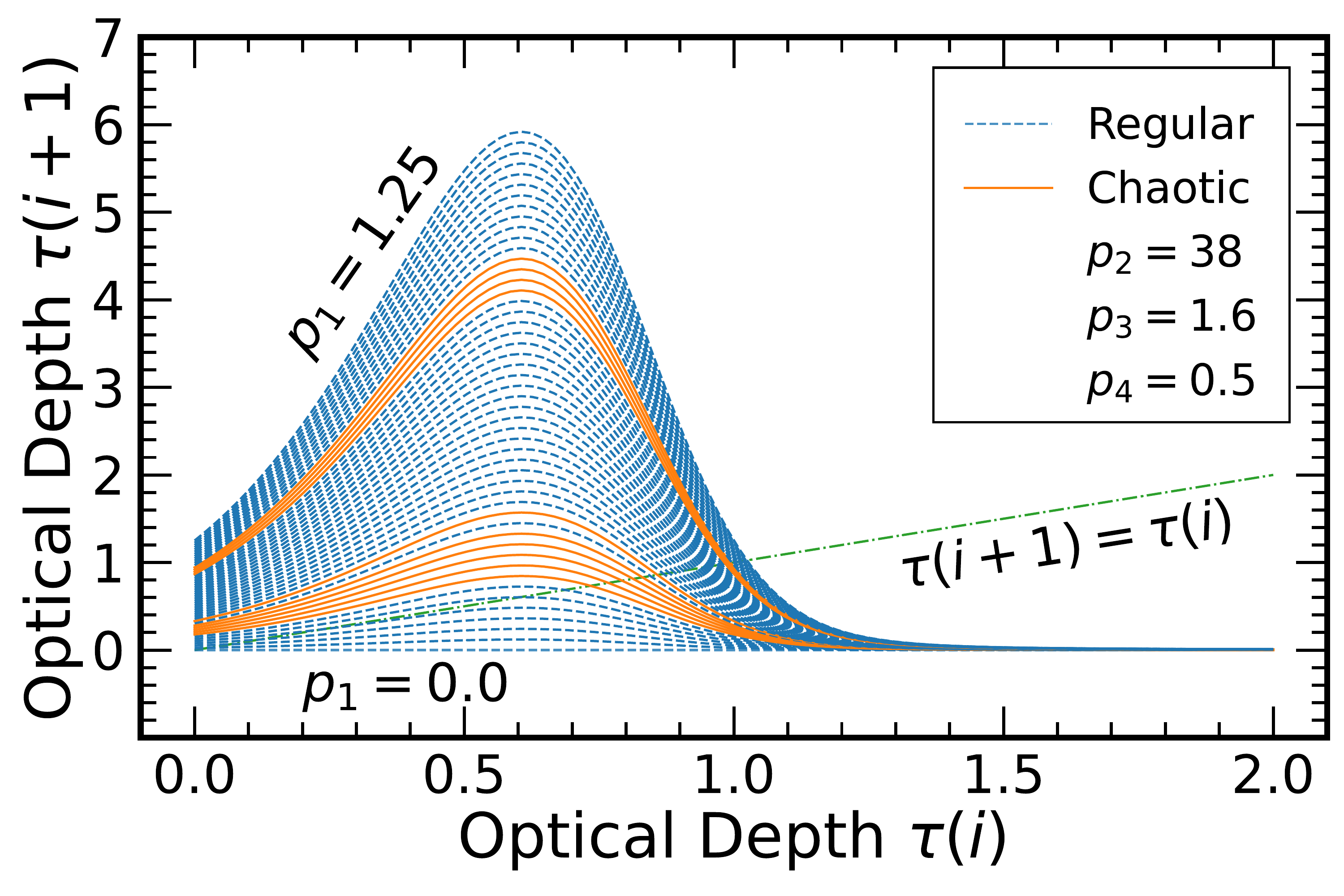}
\vspace{-0.0in}
\caption{A family of Map B iteration curves at high $p_3 = 1.6$, complementing the orbit diagram made for the same set of parameters in Fig.~\ref{fig:orbitDiagramB}. Chaotic maps are in solid orange while regular maps are in dashed blue.
%Map Family for $p_2 = 38$, $p_3 = 1.6$, $p_4 = 0.5$.
}
%Fig 12
\label{fig:mapFamilyB}
\end{figure}

\begin{figure*}
\centering 
%\vspace{-0.75in}
\includegraphics[width=0.8\textwidth]{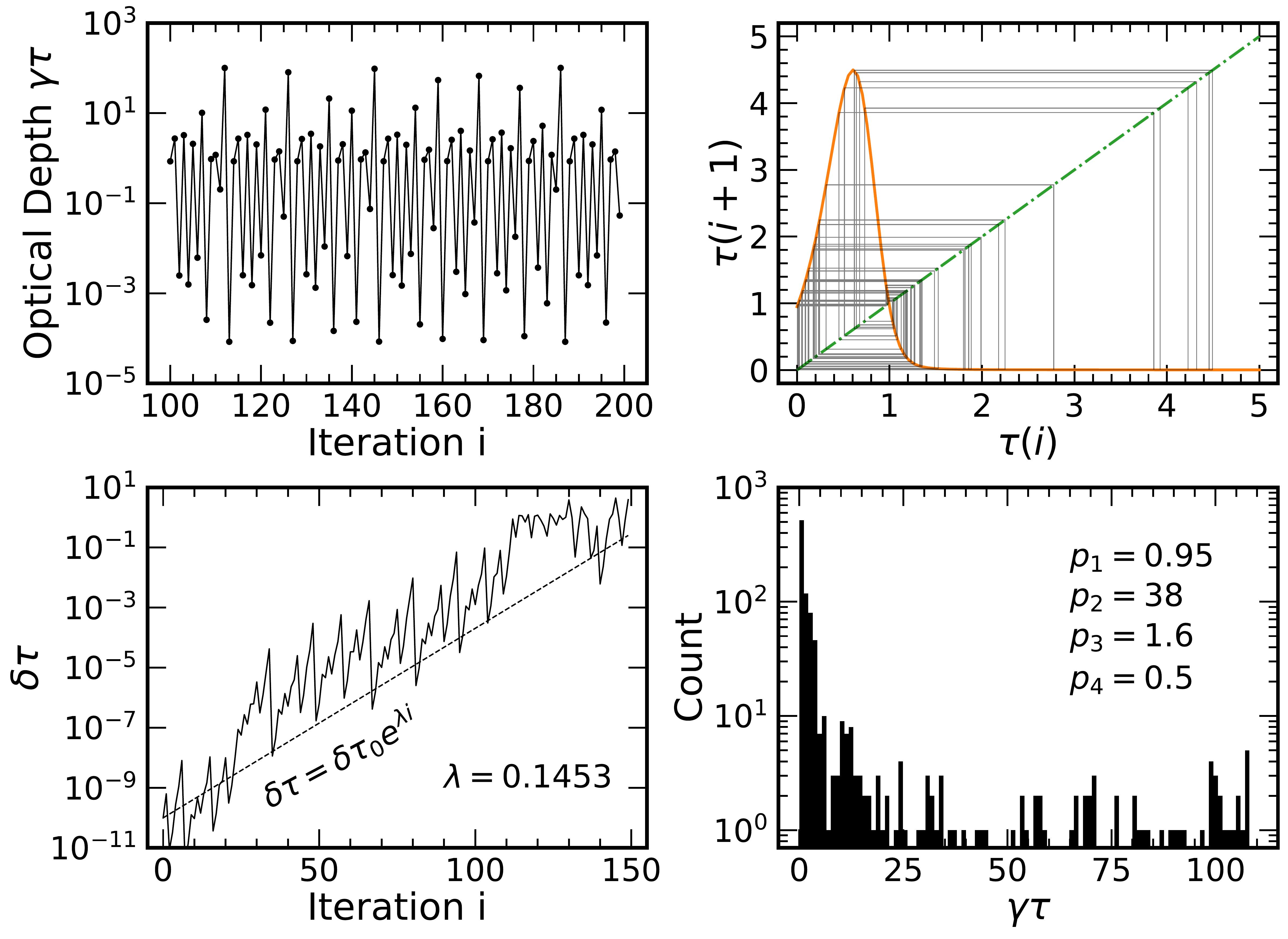}
\vspace{-0.0in}
\caption{A trajectory in the chaotic region at $p_1 \sim 0.9$ in Fig. \ref{fig:orbitDiagramB}. Because the post-clear-skies optical depth $p_1$ (the leftmost point on the iteration curve) is close to the unstable fixed point, the trajectory repeatedly returns there. Note the large dynamic range of $\gamma \tau$; recall that $\gamma$ varies between $10^{-p_3}$ and $10^{+p_3}$, and here $p_3 = 1.6$. Other map parameters are $p_1 = 0.95$, $p_2 = 38$, and $p_4 = 0.5$.}
%Fig 13
\label{fig:timeSeries9}
\end{figure*}

Chaotic regions of $\{p1, p2, p3\}$ parameter space are identified in Figure \ref{fig:locus1}. The chaos we have described above, for $0.5 \lesssim p_1 \lesssim 0.85$, $p_2 \approx 38$, and $p_3 \approx 0.6$ -- 0.75, occupies the lower lip of the `valley' seen in the bottom left panel of Fig.~\ref{fig:locus1}. If we fix $p_2 = 38$ but now allow $p_3$ to increase, the chaotic locus splits into two portions --- the two `walls' of the valley --- one at low $p_1 \sim 0.2$, and another at high $p_1 \sim 1$. An orbit diagram made at high $p_3 = 1.6$, shown in Figure \ref{fig:orbitDiagramB}, sheds further light on the two chaotic regions. The chaos at high $p_3 = 1.6$ and low $p_1 \sim 0.25$ is akin to the chaos discussed earlier at $p_3 = 0.6$ and 
$p_1 \sim 0.6$: repeated bifurcations lead to chaos, reverting back to a regular low-$n$ cycle for high enough $p_1$. This high-$p_3$, low-$p_1$ chaotic region corresponds to the lower set of chaotic (orange solid) iteration curves displayed in Figure \ref{fig:mapFamilyB}, analogous to the chaotic iteration curves of Fig.~\ref{fig:mapFamilyA}. Both $p_1$ and $p_3$ serve to increase the height of the iteration curve, and so to ensure the iteration curve is not so tall that the map reverts to a regular low-$n$ cycle, $p_1$ and $p_3$ trade off against each other to maintain this first regime of chaos.

The second chaotic region at high $p_1 \sim 0.9$ and high $p_3 = 1.6$ is new, and corresponds to the taller set of orange solid curves in Fig.~\ref{fig:mapFamilyB}. A
 sample time series for $p_1 = 0.95$ is showcased in Figure \ref{fig:timeSeries9}. In this second chaotic region, $p_1$ nearly coincides with the unstable fixed point. This near-coincidence implies that after every clear sky with $\tau \ll 1$, the system subsequently lands near the fixed point.
 %After every clear sky with $\tau(i) \ll 1$, the dusty wind blows anew, and one iteration later the atmosphere's optical depth attains nearly its equilibrium value, $\tau(i+1) \simeq p_1 \simeq \tau_{\rm f}$ (where the solid orange and green dot-dashed curves intersect). 
Because the point is unstable, the system spirals away from it, exploring a few other values of $\tau$ before landing on the map's nuclear-winter tail at high $\tau \gtrsim 1.5$.
%The closer to the fixed point the system lands, the longer it takes to spiral away, and the more values of $\tau$ are sampled. 
%Eventually the system spirals to high $\tau$ on the map's nuclear-winter tail. 
The system then re-sets to a
nearly clear sky with $\tau \ll 1$; upon re-starting, it now lands slightly differently relative to the unstable point
and thereby generates a different $\tau$ sequence. Thus the
system explores a continuum of $\tau$ values. 

We conclude our study of Map B by assessing the effect of varying $p_2$. The bottom right panel of Fig.~\ref{fig:locus1} shows that decreasing $p_2$ increases the values of $p_3$ required for chaos; the bottom of the valley, tracing the first regime of chaos, slopes upward from high $p_2$/low $p_3$, to low $p_2$/high $p_3$. This trend is sensible since lowering $p_2$ decreases how much the map varies from iteration to iteration --- see equation (\ref{map}) where $p_2$ enters in the exponent with $\gamma\tau(i)$. To offset the loss of sensitivity from lowering $p_2$, the parameter $p_3$, which controls the magnitude of $\gamma$, must increase.

\subsection{Map C: $\gamma(\tau)$ and Above-Ground Temperature} \label{subsec:C}

Equation (\ref{Tg}) gives the ground temperature $T$ in a 2-stream radiative equilibrium model (\citealt{pierrehumbert10}), for spatially constant $\gamma \equiv \kappa_{\rm V}/\kappa_{\rm IR}$ and total atmospheric $\tau$ (= $\tau_\infty$ in Pierrehumbert's notation). In this solution, the ground temperature $T$ formally differs from the atmospheric temperature just above the ground, $T'$:
\begin{align}
T'(i) &= c_3 [  (1+1/\gamma) + (\gamma-1/\gamma)e^{-\gamma\tau(i)}]^{1/4} \label{Tg2}
\end{align}
where we see that the $(1-1/\gamma)$ term in equation (\ref{Tg}) has been replaced by $(\gamma-1/\gamma)$. The spatially discontinuous jump from $T$ to $T'$ in the 2-stream radiative solution is unphysical; in reality, it is smoothed away by a combination of conduction between the ground and the atmosphere, and convection. Rather than add this extra physics to our model, we take $T'$ to represent a different limiting case for the ground temperature, using it instead of $T$ in equations (\ref{mdot}), (\ref{tau_mdot}), and (\ref{gamma}) to create a new Map C:
\begin{align} \label{mapC}
\tau(i+1) = & \,\, p_1 
%\exp( +2^{-1/4} p_2 ) 
%\exp[ +(1+\gamma)^{-1/4} p_2] \,
\exp[ +(1+10^{-p_3})^{-1/4} p_2] \,
\times \\
& \exp \{ -p_2 [(1+1/\gamma) + (\gamma-1/\gamma) \exp(-\gamma\tau(i))]^{-1/4} \}  \nonumber
\end{align}
where the constant pre-factor depending on $p_1$, $p_2$, and $p_3$ is such that $p_1$ is still interpretable as the post-clear-sky optical depth (i.e. if $\tau(i) \ll 1$, then $\tau(i+1) = p_1$). As with Map B, $\gamma = \gamma(i)$ depends on $\tau(i)$ through equation (\ref{gamma}). Our goal in exploring this new Map C is to get a sense of how sensitive outcomes are to model details.

Fig.~\ref{fig:Tg} shows how the new ground temperature $T'$ varies with the total visible optical depth $\gamma \tau$ of the atmosphere. The variation is non-monotonic, more so than for Map B, ensuring that Map C is also non-invertible. When the atmosphere is optically thin, $T'$ is lower than $T$ because, in the context of the 2-stream solution,  the atmosphere is less easily heated than the ground. Under optically thick conditions, Maps B and C converge to a common low temperature.

\begin{figure*}
\centering 
%\vspace{-0.75in}
\includegraphics[width=0.8\linewidth]{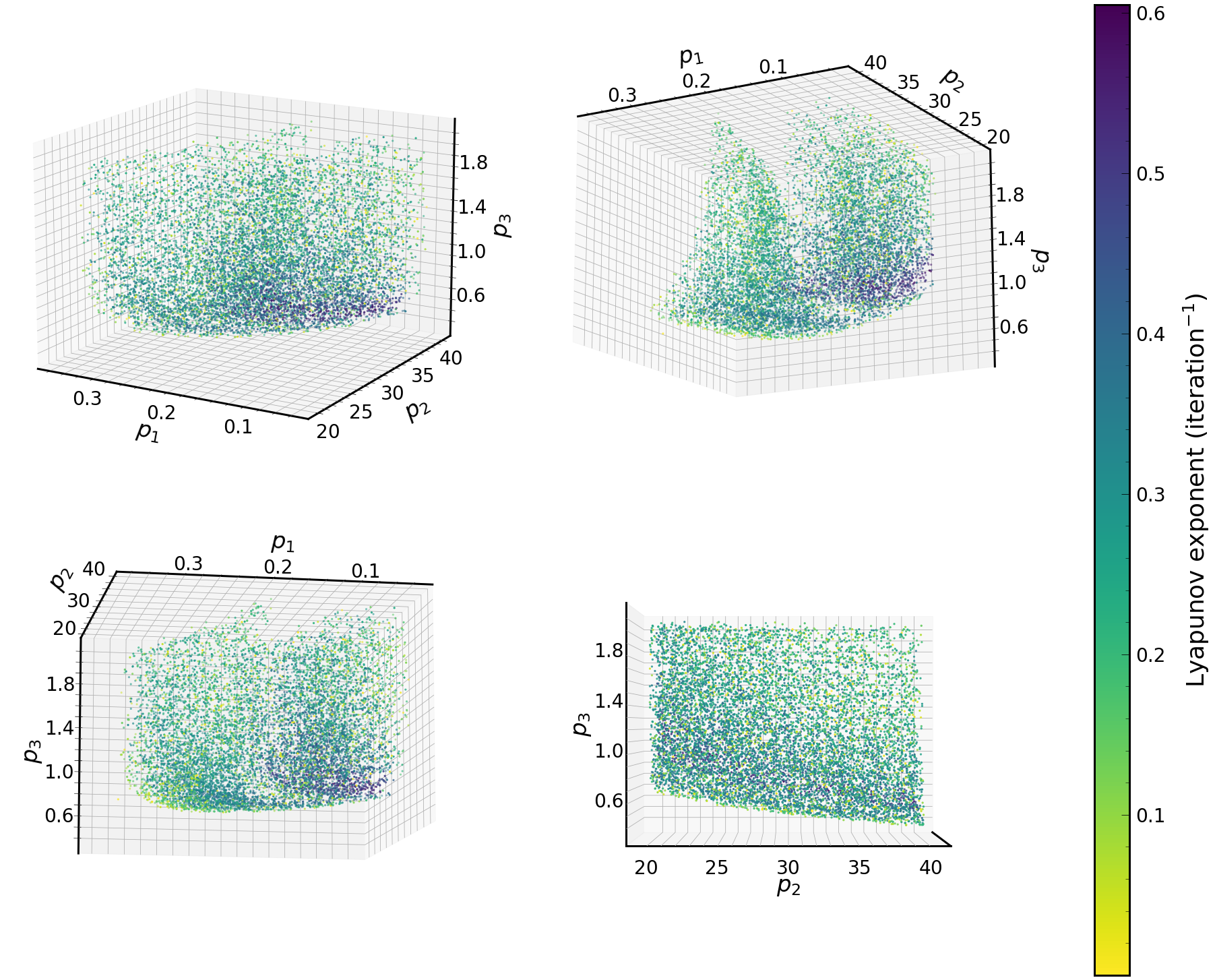}
\vspace{-0.0in}
\caption{Chaotic locus of Map C. At each of 50000 points randomly
  distributed in the 3D space of $0 < p_1 < 0.4$, $20 < p_2 < 40$ and $0 < p_3 <
  2$ ($p_4 =0.5$ is fixed), the Lyapunov exponent is calculated; if positive, that point is
  marked on the plot, colored according to the magnitude of the
  exponent. All panels show the same data viewed from different angles around the $p_3$ axis.}
%Fig 14
\label{fig:locus2}
\end{figure*}

\begin{figure}
\centering
\includegraphics[width=1.0\linewidth]{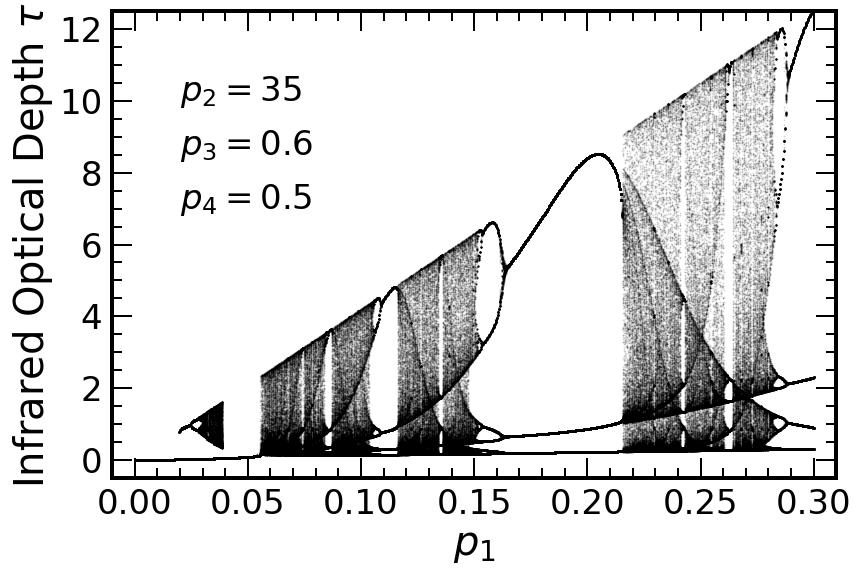}
\vspace{-0.0in}
\caption{Orbit diagram for Map C. To construct this plot, we iterate Map C 200 times for a given $p_1$ starting from each of 5 values of $\tau(i=0)$ spaced evenly from 0 to 12 inclusive and plot the last 100 values of $\tau(i)$. For $0.02 \lesssim p_1 \lesssim 0.04$, depending on the initial condition, the trajectory either converges to an equilibrium or is chaotic. At $p_1 \approx 0.06$, the trajectory switches from a single-valued equilibrium to chaos; contrast this with the series of bifurcations leading to chaos in Map B. See also Fig. \ref{fig:multipleFixed}. Map parameters are $p_2 = 35$, $p_3 = 0.6$ and $p_4 = 0.5$.}
%Fig 15
\label{fig:orbitDiagramC}
\end{figure}

In Figure \ref{fig:locus2} we identify the regions of $\{p_1,p_2,p_3\}$ space that generate chaos, fixing $p_4 = 0.5$ as usual. The chaotic locus for Map C appears to have roughly the same U-shape as for Map B (Fig.~\ref{fig:locus1})---there is a branch at low $p_1$/high $p_3$, and another branch at high $p_1$/high $p_3$. The orbit diagram for Map C in Figure \ref{fig:orbitDiagramC}, showing the two chaotic regimes separated by a regular window at $0.15 \lesssim p_1 \lesssim 0.21$, also echoes its Map B counterpart in Fig.~\ref{fig:orbitDiagramB}.

\begin{figure*}
\centering 
%\vspace{-0.75in}
\includegraphics[width=0.8\textwidth]{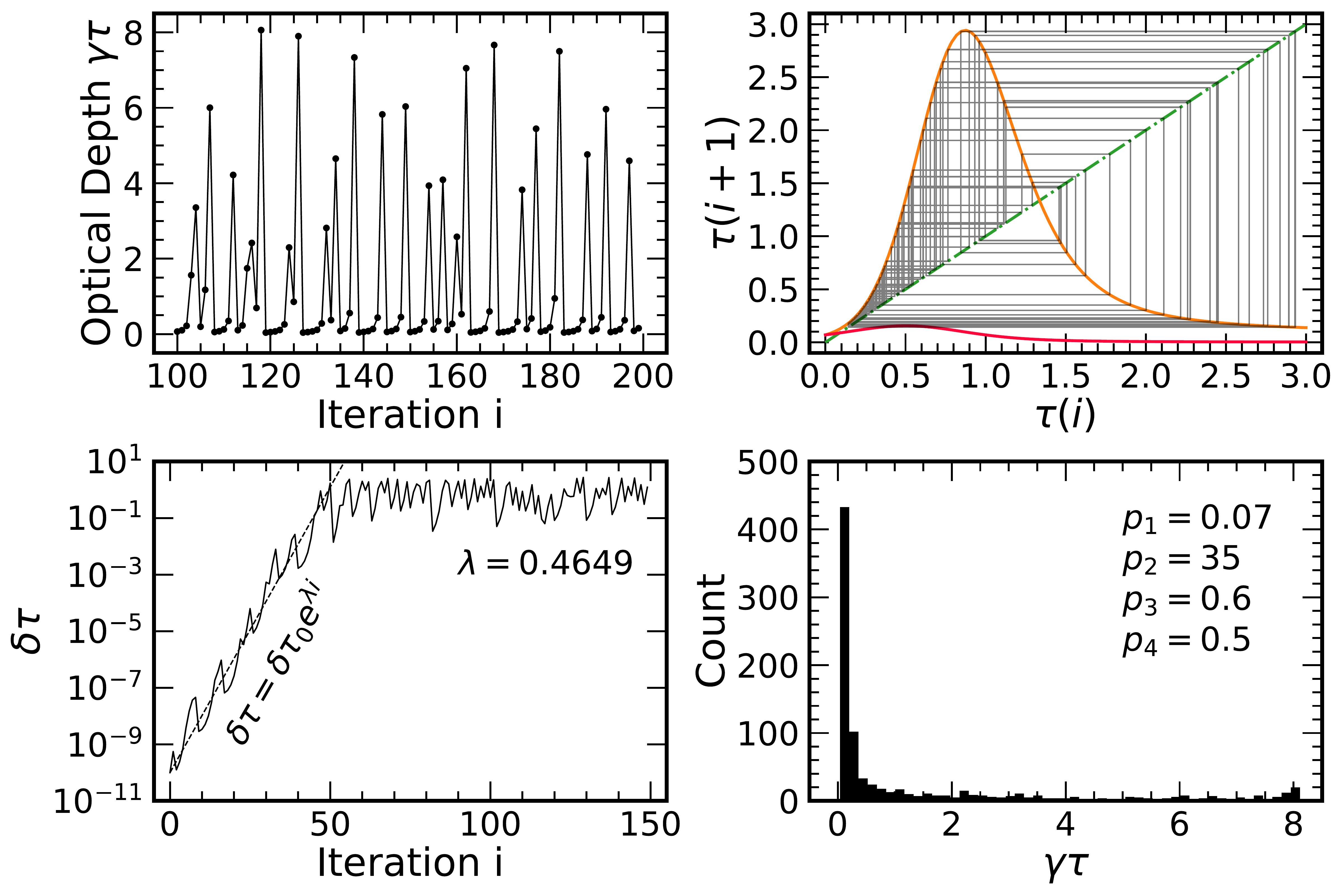}
\vspace{-0.0in}
\caption{A Map C trajectory. Notice how $\gamma\tau$ builds up over several iterations before each spike. Chaos is possible here at small $p_1= 0.07$, unlike for Map B, because for the same parameters $\{p_1, p_2, p_3, p_4\}$, the iteration curve for Map C (orange) is higher than that for Map B (red), leading to an unstable fixed point for the former.  Other map parameters are $p_2 = 35$, $p_3 = 0.6$, and $p_4 = 0.5$.}
%Fig 16
\label{fig:timeSeriesC}
\end{figure*}

One quantitative difference between the two maps
is that the values of $p_1 \lesssim 0.3$ which lead to chaos in Map C are for the most part lower than for Map B. An example comparison between the two maps in Figure \ref{fig:timeSeriesC}, made at fixed $p_1 = 0.07$, $p_2 = 35$, and $p_3 = 0.6$, illustrates why. At this common low value of $p_1$, Map C is chaotic while Map B is not: the maximum value of $\tau(i+1)$ attained by Map C is larger than for Map B, and the resultant steeper slope of Map C's iteration curve de-stabilizes its fixed point. The larger maximum $\tau(i+1)$ stems from $c_1 c_4$ being larger for Map C than for Map B for common $\{p_1,p_2,p_3\}$; for Map C, $p_1 = c_1c_4 \exp [- (1+10^{-p_3})^{-1/4}p_2]$, whereas for Map B, $p_1 = c_1c_4 \exp [-2^{-1/4}p_2]$. Larger $c_1c_4$ implies a planet that emits a stronger, dustier wind, all other factors being equal (eqs.~\ref{mdot} and \ref{tau_mdot}). Increasing $c_1c_4$ is how Map C can achieve the same post-clear-skies optical depth $p_1$ as Map B in the face of Map C's cooler clear-skies surface temperature.

\begin{figure*}
\centering 
%\vspace{-0.75in}
\includegraphics[width=0.8\textwidth]{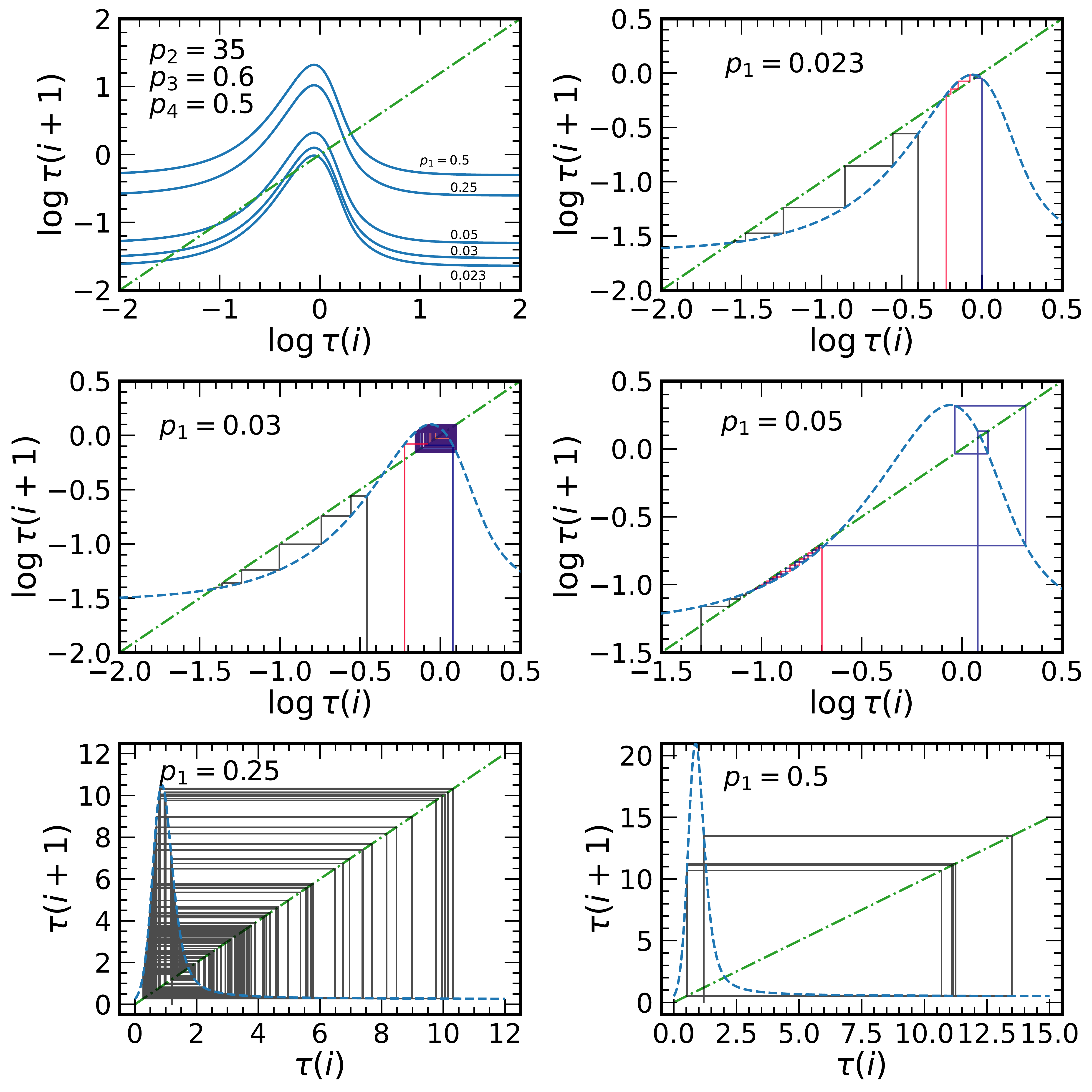}
\vspace{-0.0in}
\caption{Cobwebs for Map C. The iteration curves in the top left panel are displayed individually in the other panels. Unlike Map B, Map C can exhibit multiple fixed points, with different initial conditions leading to orbits around different points. Three trajectories with different initial conditions are shown in blue, red, and black for each of the $p_1 = (0.023,0.02,0.05)$ panels. Widespread chaos is seen for $p_1 = 0.25$ but disappears for still higher $p_1$.}
%\textit{Top left:} Comparison between Map C iteration curves for differing values of $p_1$. Map parameters are $p_2 = 35$, $p_3 = 0.6$, $p_4 = 0.5$. \textit{Top right:} Cobweb plot for $p_1 = 0.023$. The trajectory in black converges to the lowest fixed point and the trajectories in red and blue converge to the highest fixed point. \textit{Middle left:} Cobweb plot for $p_1 = 0.03$. The trajectory in black converges to the lowest fixed point and the trajectories in red and blue chaotically orbit the highest fixed point. \textit{Middle right:} Cobweb plot for $p_1 = 0.05$. All three trajectories converge to the lowest fixed point. \textit{Bottom left: } Cobweb plot for $p_1 = 0.25$ illustrating a chaotic trajectory for a larger value of $p_1$. \textit{Bottom right:} Cobweb plot for $p_1 = 0.5$ illustrating a regular trajectory for a larger value of $p_1$}
%Fig 17
\label{fig:multipleFixed}
\end{figure*}

Map C also differs from Map B in allowing for the possibility of multiple fixed points, as illustrated in Fig.~\ref{fig:multipleFixed}. At small $p_1 \lesssim 0.05$, the iteration curve intersects the $\tau(i+1) =\tau(i)$ line at three locations, a consequence of Map C's $T'(\tau)$ relation being more strongly non-monotonic (Fig.~\ref{fig:Tg}). Among the three fixed points there are both stable and unstable points, and whether a trajectory is chaotic or regular, and which of the stable points a regular trajectory converges to, depend on initial conditions. These complications are also reflected in the orbit diagram of Fig.~\ref{fig:orbitDiagramC} at small $p_1$, which does not show the comparatively simple bifurcation sequence of Map B. 
The particular trajectory shown in Fig.~\ref{fig:timeSeriesC} for $p_1
= 0.07$ tends to linger 
%near $\tau \sim 0.2$
at low visible optical depth $\gamma \tau \ll 1$ for several
iterations at a time, a consequence of there `almost' being a fixed
point at $\tau \sim 0.2$ (the fixed points for $p_1 < 0.05$ become
just unfixed at $p_1 > 0.05$; Fig.~\ref{fig:multipleFixed}). This lingering at low optical depth is reminiscent of, but shorter in duration than, the `off' phases exhibited by KIC 1255b when no transit is detectable for up to dozens of orbital periods 
(\citealt{rappaport12,vanwerkhoven14}).

Apart from these differences, however, Maps C and B exhibit similar behaviours. In the limit of large $p_1$, peak $\tau(i+1)$ values are so high that the evolution reduces to a simple, regular boom-bust cycle. Trends with $p_2$ and $p_3$ are quantitatively similar between the two maps. Chaos prevails for $20 \lesssim p_2 \lesssim 40$ and $p_3 \gtrsim 0.6$, with higher $p_2$ (greater sensitivity of mass-loss rate to optical depth) preferred at lower $p_3$.

\section{Summary and Discussion}\label{sec:sum}

Planets too close to their host stars vaporize. As observed by {\it Kepler}, a rocky planet in its final evaporative throes loses mass in fits and starts.

We have demonstrated with a simple one-dimensional map that an evaporative wind can alternately boom and bust in a regular 2-cycle if:
\begin{enumerate}
\item The wind mass-loss rate rises exponentially with planet surface temperature (\`a la Clausius-Clapeyron);
\item The instantaneous optical depth to the wind scales with the wind mass-loss rate at a prior time (there is hysteresis from the wind's finite speed); 
\item The planet surface temperature decreases with increasing optical depth of the wind (as in a radiative equilibrium atmosphere whose opacity to incoming starlight is greater than its opacity to reprocessed infrared radiation). 
\end{enumerate}
The period of the boom-bust cycle is the atmospheric
refresh time --- how long the wind takes to fill 
however much of 
the
space 
it can 
between the planet's dayside surface and the star. 
This time, defined as $t_{\rm travel}$ in section \ref{sec:intro}, is comparable to the orbital period, because that is how long it
takes a thermal wind to travel out to the planet's Hill radius (\citealt{perezbecker13}). An `on-off' behaviour with deep and shallow transits alternating with 
every orbit has been observed for KIC 1255b for
stretches lasting on the order of $\sim$10 orbits \citep{vanwerkhoven14}.

Most of the time, however, the transit depths of KIC 1255b betray no pattern. We have found that the 2-cycle of our one-dimensional map erupts into a chaotic high-$n$-cycle if we revise condition (iii) above to read:
\begin{enumerate}
\setcounter{enumi}{2}
\item (Revised) The planet surface temperature increases with increasing optical depth when the atmosphere is optically thin, and decreases with increasing optical depth when the atmosphere is optically thick.
\end{enumerate}
There is an {\it a priori} reason for the revised condition (iii). Under optically thin, clear sky conditions, the only dust grains that can condense out of the evaporative wind must have visible wavelength opacities lower than their infrared wavelength opacities. An opacity ratio $\gamma < 1$ is needed for dust grains to radiate away their energy more efficiently than they absorb visible-wavelength starlight; otherwise grains would be heated to super-blackbody temperatures exceeding that of the vaporizing planet surface, and would not be able to condense in the first place. Condensates with the right opacity ratio are iron-poor and silicate-rich (over a wide range of grain sizes, from sub-micron to super-micron; see figure 4 of \citealt{booth23}), and they induce a greenhouse effect which raises the planet's surface temperature by infrared back-warming --- this satisfies the first clause of condition (iii). Once enough grains condense to render the atmosphere optically thick to starlight, they can acquire iron and absorb more efficiently in the visible than the infrared, as conventional micron and sub-micron grains do. The reversed opacity ratio $\gamma > 1$ under optically thick conditions induces a nuclear winter that cools the planetary surface, satisfying the second clause of condition (iii).

Chaos results from our one-dimensional map provided the magnitude of the opacity ratio `flip' from $\gamma <1$ to $\gamma >1$ is at least a factor of 10. A flip of this magnitude is plausible between iron-poor and iron-rich silicates, although for iron-poor silicates $\gamma$ is not much below 1 (J.~Owen, personal communication 2023). Photon scattering, which we have neglected, could be significant at stellar optical wavelengths if grains grow past $\sim$0.1 $\mu$m in size (in the opposing Rayleigh limit, when grain sizes are small compared to the wavelength, scattering cross sections are lower than absorption cross sections, even for a poorly absorbing, iron-poor mineral like quartz). But large grains are hard for the wind to lift (e.g.~\citealt{perezbecker13}), and for plane-parallel atmospheres, scattering attenuates incident radiation more weakly than exponentially \citep[section 4.1 of][]{chamberlainhunten87}.

Another requirement for chaos is that the infrared optical depth one atmospheric refresh time after a clear sky be somewhere between 0.05 and 1, the exact interval depending on other parameters. Too low a post-clear-sky optical depth, and the wind settles into an optically thin steady state. Too high a post-clear-sky optical depth, and the wind locks into a regular boom-bust 2-cycle. 
%What conditions prevail in reality are unclear and require more detailed physical models of the kind developed by, e.g., Booth et al.~(2022).

Condition (iii) relates planet surface temperature to stellar irradiance and the radiative properties of the overlying atmosphere, and contains a number of assumptions. Evaporative cooling is ignored, as is heat transport from the planet's surface into its interior. The first effect is indeed negligible, and so is the second if transport is by thermal conduction. The cooling flux from vaporization is $L_{\rm vap}\dot{M}/R^2 \sim 0.4$ kW/m$^2$, where $L_{\rm vap} \sim 10^{11}$ erg/g is the latent heat of vaporization of rock/iron, $\dot{M} \sim 1 M_\oplus/{\rm Gyr}$ is the planet mass-loss rate, and $R \sim R_\oplus/3$ is the planet radius (values drawn from \citealt{perezbecker13}). This evaporative energy flux is tiny compared to the incoming stellar flux of $\sim$$10^3$ kW/m$^2$. The conductive heat flux is also a small perturbation. Over an atmospheric refresh time of $t_{\rm travel} \sim$ 10 hr, heat can diffuse across a crustal thickness of $\Delta \ell \sim \sqrt{\kappa t_{\rm travel}} \sim 20$ cm, where $\kappa \sim 10^{-2}$ cm$^2$/s is the thermal diffusivity of rock. The conductive heat flux across such a layer,  having thermal conductivity $k_{\rm c} \sim 4$ W/m/K, is at most $k_{\rm c} \Delta T / \Delta \ell \sim 40$ kW/m$^2$, an upper limit calculated using a temperature difference of $\Delta T \sim 2000$ K across the crust (the actual $\Delta T$ would  almost certainly be lower). A liquid magma ocean that convects rather than conducts heat might change this calculus, depending on the depth of the ocean and the speed of convective eddies. 

Condition (iii) further assumes the wind achieves radiative equilibrium on a timescale fast compared to the refresh time $t_{\rm travel} \sim 10$ hr. The radiative thermal time is $t_{\rm thermal} \sim \Sigma k (\tau + 1/\tau)/ (\mu m_{\rm H} \sigma T^3 )$, where $\Sigma = \tau/\kappa$ is the mass column density, $\tau$ is the infrared optical depth, $\kappa$ is the opacity, $\mu$ is the mean molecular weight, $k$ is Boltzmann's constant, $\sigma$ is the Stefan-Boltzmann constant, and $m_{\rm H}$ is the mass of hydrogen. For approximately constant $T \sim 2000$ K, the thermal time attains its minimum when $\tau \lesssim 1$; we estimate $\min t_{\rm thermal} \sim 1$ s for $\kappa \sim 20$ cm$^2$/g (this opacity derives from the right panel of figure A1 of \citealt{booth23}; see their `Dust,  $\times \, 0.01$' curve which assumes a dust-to-gas mass ratio of 0.01). Thus $t_{\rm thermal} \ll t_{\rm travel}$ for $\tau \lesssim 30$.

Chaotic winds in our maps have Lyapunov times ranging from $\sim$2 to $\sim$10 atmospheric refresh times. The probability distribution of visible optical depths in a given map typically features a peak at zero and a tail that extends to optically thick values (3--10, or greater, depending on map parameters). The peak at zero reflects the system in its `off' state --- even when the wind is chaotic, it still retains aspects of a 2-cycle, alternating between boom and bust phases every one or few refresh times. Only by modeling the cometary tail emitted by a disintegrating planet can we properly relate the wind optical depths from our maps to actual transit depths. That said, the shapes of our optical depth distributions, which also sometimes exhibit gaps and multiple modes, do not obviously match the shape of the distribution of visible-wavelength transit depths measured for KIC 1255b. The latter distribution does not peak at zero, but at a finite transit depth corresponding to a wind that is probably marginally optically thick; the transit depth distribution falls smoothly and nearly symmetrically to either side of this peak (see figure 4 of \citealt{vanwerkhoven14}). Also remaining to be explained are the quiescent intervals lasting up to 36 days $\sim$ 50 atmospheric refresh times, when KIC 1255b displays no detectable transits (\citealt{rappaport12,vanwerkhoven14,schlawin18}). By contrast, the longest our map lingers at low optical depth is $\sim$5 refresh times.

There are assuredly sources of randomness not captured by our minimalist map. On our suspect list are the possibility that a time-variable stellar wind can shape the planetary wind and tail (cf.~\citealt{kawahara13, croll15, schlawin18}), analogous to how the magnetized Solar wind can sculpt comet ion and dust tails \citep{ip04,price19,price23};
%particularly its extension into the comet-like tail that controls the transit depth (e.g.~Ridden-Harper et al.~2018),
%(Rappaport et al.~2012; Brogi et al.~2012; Budaj 2013; van Lieshout et al.~2014), 
the vagaries of dust nucleation (a.k.a.~cloud formation); and
changes in planet surface albedo due to fallback and removal of dust. Global dust storms on Mars, which occur stochastically on timescales of $\sim$1--10 Martian years, might also provide relevant insights into positive and negative feedbacks between dust and atmospheric heating, and how dust is transported across the planet surface by horizontal winds \citep{kahre17}.

\section*{Acknowledgements}
We thank Edgar Knobloch for fostering this collaboration, Saul Rappaport for encouraging exchanges, Nick Choksi for assistance with making figures, and the Berkeley Physics-and-Astrophysics Undergraduate Research Stipend (BPURS) for financial support. Edwin Kite  provided an extensive and thoughtful review of a draft version of this paper that led to substantive improvements in presentation, and checks on our neglect of conductive and evaporative energy fluxes. Constructive reviews were also given by Richard Booth, James Owen, and an anonymous referee. This work also benefited from an airing at the Penn State Exoplanet Journal Club. Our running title and the phrase ``does not go gentle into that good night'' used in the introduction are taken from Dylan Thomas's poem, ``Do not go gentle into that good night'', held in copyright by the Dylan Thomas Trust.

%%%%%%%%%%%%%%%%%%%%%%%%%%%%%%%%%%%%%%%%%%%%%%%%%%
\section*{Data Availability} 
Data and codes for generating our figures are available upon request of the authors. The code for generating the orbit diagram in Figure \ref{fig:orbitDiagramA} is available at \url{https://github.com/joshuaabromley/bromley_chiang_map}.

%The inclusion of a Data Availability Statement is a requirement for articles published in MNRAS. Data Availability Statements provide a standardised format for readers to understand the availability of data underlying the research results described in the article. The statement may refer to original data generated in the course of the study or to third-party data analysed in the article. The statement should describe and provide means of access, where possible, by linking to the data or providing the required accession numbers for the relevant databases or DOIs.



\bibliographystyle{mnras}
\bibliography{GI} % if your bibtex file is called example.bib

% End of mnras_template.tex

%\begin{thebibliography}{99}
%\end{thebibliography}

% Don't change these lines
\bsp	% typesetting comment
\label{lastpage}
\end{document}